\documentclass[aps,prd,10pt,twocolumn,preprintnumbers,nofootinbib,superscriptaddress,longbibliography]{revtex4-2}

\usepackage{graphicx}
\usepackage{amsmath, amssymb}
\usepackage{hyperref}
\usepackage{booktabs}
\hypersetup{
  colorlinks  = true,
  urlcolor    = blue,
  linkcolor   = blue,
  citecolor   = red
}
\usepackage[capitalize]{cleveref}

\newcommand{\orcid}[1]{\begingroup
  \hypersetup{hidelinks}\href{https://orcid.org/#1}{\includegraphics[width=10pt]{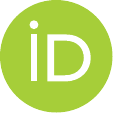}} \endgroup}

\usepackage{xcolor}
\definecolor{Thispaper}{HTML}{ff5f20}
\definecolor{Planck18}{HTML}{2166ac}

\begin{document}

\title{Origin of cosmological neutrino mass bounds: background \emph{versus} perturbations}

\author{Toni Bertólez-Martínez\orcid{0000-0002-4586-6508}}
\email{antoni.bertolez@fqa.ub.edu}
\affiliation{Departament de Física Quàntica i Astrofísica and Institut de Ciències del Cosmos, Universitat de Barcelona, Diagonal 647, E-08028 Barcelona, Spain}

\author{Ivan Esteban\orcid{0000-0001-5265-2404}}
\email{ivan.esteban@ehu.eus}
\affiliation{Department of Physics, University of the Basque Country UPV/EHU, PO Box 644, 48080 Bilbao, Spain}
\affiliation{EHU Quantum Center, University of the Basque Country UPV/EHU}

\author{Rasmi Hajjar\orcid{0000-0002-9227-5364}}
\email{rasmi.hajjar@ific.uv.es}
\affiliation{Instituto de F\'{i}sica Corpuscular (IFIC), University of Valencia-CSIC, Parc Cient\'{i}fic UV, c/ Cate\-dr\'{a}tico Jos\'{e} Beltr\'{a}n 2, E-46980 Paterna, Spain}

\author{Olga Mena\orcid{0000-0001-5225-975X}}
\email{omena@ific.uv.es}
\affiliation{Instituto de F\'{i}sica Corpuscular (IFIC), University of Valencia-CSIC, Parc Cient\'{i}fic UV, c/ Cate\-dr\'{a}tico Jos\'{e} Beltr\'{a}n 2, E-46980 Paterna, Spain}

\author{Jordi Salvado\orcid{0000-0002-7847-2142}}
\email{jsalvado@fqa.ub.edu}
\affiliation{Departament de Física Quàntica i Astrofísica and Institut de Ciències del Cosmos, Universitat de Barcelona, Diagonal 647, E-08028 Barcelona, Spain}

\date{\today}
\preprint{}

\begin{abstract}
The cosmological upper bound on the total neutrino mass is the dominant limit on this fundamental parameter. Recent observations---soon to be improved---have strongly tightened it, approaching the lower limit set by oscillation data. Understanding its physical origin, robustness, and model-independence becomes pressing. Here, we explicitly separate for the first time the two distinct cosmological neutrino-mass effects: the impact on background evolution, related to the energy in neutrino masses; and the ``kinematic'' impact on perturbations, related to neutrino free-streaming. We scrutinize how they affect CMB anisotropies, introducing two effective masses enclosing \emph{background} ($\sum m_\nu^\mathrm{Backg.}$) and  \emph{perturbations} ($\sum m_\nu^\mathrm{Pert.}$) effects. We analyze CMB data, finding that the neutrino-mass bound is mostly a background measurement, i.e., how the neutrino energy density evolves with time. The bound on the ``kinematic'' variable $\sum m_\nu^\mathrm{Pert.}$ is largely relaxed, $\sum m_\nu^\mathrm{Pert.} < 0.8\,\mathrm{eV}$. This work thus adds clarity to the physical origin of the cosmological neutrino-mass bound, which is mostly a measurement of the neutrino equation of state, providing also hints to evade such a bound.
\end{abstract}

\maketitle


\section{Introduction}
\label{sec:intro}
Neutrinos are hot thermal relics, permeating the Universe and being the most abundant particles after Cosmic Microwave Background (CMB) photons. In particle physics, neutrino masses have meant the first departure from the Standard Model (SM), where no gauge-invariant renormalizable neutrino mass term can be written. Yet neutrino oscillation experiments have robustly measured two squared-mass differences, $|\Delta m^2_{31}|\equiv|m^2_3-m^2_1|\simeq 2.5\cdot 10^{-3}\,\mathrm{eV}^2$ and $\Delta m^2_{21} \equiv m^2_2-m^2_1 \simeq 7.5\cdot 10^{-5}\,\mathrm{eV}^2$~\cite{deSalas:2020pgw, Esteban:2024eli, Capozzi:2021fjo}, confirming that neutrinos are massive and that the SM must be extended. Two mass orderings are possible, \emph{normal} (NO, $\Delta m^2_{31}>0$) and \emph{inverted} (IO, $\Delta m^2_{31}<0$), and oscillation experiments aim to determine it~\cite{T2K:2023smv, NOvA:2021nfi, JUNO:2015zny, Hyper-Kamiokande:2018ofw, DUNE:2020ypp}. 

The absolute neutrino mass scale, however, remains unknown, as oscillations are only sensitive to mass differences. Oscillation results constrain $\sum m_\nu > 0.06\,\mathrm{eV}$ for NO and $\sum m_\nu > 0.1\,\mathrm{eV}$ for IO~\cite{Esteban:2024eli}, and direct searches in the KATRIN experiment imply $\sum m_\nu < 1.35\,\mathrm{eV}$~\cite{Katrin:2024tvg}; where $\sum m_\nu \equiv m_1 + m_2 + m_3$ is the total neutrino mass. Even if neutrino masses are tiny, determining them is a mandatory first step to understand not only the origin of the particle masses and hierarchies, but also the Universe's evolution and structure formation---neutrinos are, so far, the only known form of (hot) dark matter. 

In the early Universe, neutrinos are relativistic and behave as radiation, constituting up to $40\%$ of the total energy density. At temperatures $\lesssim \mathrm{MeV}$, their weak-interaction rate with the primordial plasma falls below the expansion rate of the Universe. Then, they decouple and stream freely, with their momentum being redshifted by the expansion. When neutrino momenta fall below their mass, neutrinos move at non-relativistic speeds, contributing to the Universe's matter energy density. 

Due to their abundance, neutrinos leave a measurable imprint on cosmological observables. This leads to upper bounds on the total neutrino mass that are around $\sum m_\nu \lesssim 0.05$--$0.3$\,eV~\cite{DESI:2024mwx, DESI:2024hhd, Jiang:2024viw, Wang:2024hen, RoyChoudhury:2024wri}, potentially raising a tension between oscillation results and cosmological observations. As cosmological datasets grow, the neutrino mass will soon be either measured or a bound in clear conflict with oscillation experiments will be placed~\cite{Euclid:2024imf, DESI:2016fyo, SimonsObservatory:2018koc, LiteBIRD:2022cnt}. 

Cosmological inferences, however, are indirect. The cosmological neutrino-mass bound is known to be correlated with, among others, the equation of state of dark energy~\cite{Hannestad:2005gj, Yang:2020ope, Vagnozzi:2018jhn, RoyChoudhury:2018vnm, diValentino:2022njd, diValentino:2022njd, Zhao:2016ecj, Zhang:2015uhk, Guo:2018gyo, RoyChoudhury:2019hls, Li:2012vn, Li:2012spm, Zhang:2014nta, Zhang:2015rha, Geng:2015haa, Chen:2015oga, Loureiro:2018pdz, Wang:2016tsz, Yang:2017amu, Huang:2015wrx, Sharma:2022ifr, Zhang:2020mox, Khalifeh:2021ree, Chen:2016eyp, RoyChoudhury:2018vnm}; the Hubble constant $H_0$~\cite{Planck:2018vyg, Giusarma:2012ph, Hu:2023jqc, Schoneberg:2021qvd} and the amplitude parameter $\sigma_8$~\cite{Planck:2018vyg, Mccarthy:2017yqf, Abazajian:2019ejt}, that are in tension among different datasets~\cite{Abdalla:2022yfr, DiValentino:2021izs, Verde:2019ivm, Verde:2023lmm, DiValentino:2020vvd, RoyChoudhury:2018gay}; CMB lensing~\cite{RoyChoudhury:2019hls, DiValentino:2021imh, Giare:2023aix, Craig:2024tky, Green:2024xbb, Loverde:2024nfi, Naredo-Tuero:2024sgf, Allali:2024aiv}, where some observations are anomalous~\cite{Planck:2018vyg, Motloch:2018pjy, Tristram:2023haj, Rosenberg:2022sdy}; or new physics in the neutrino sector~\cite{Esteban:2021ozz, Esteban:2022rjk, Kreisch:2019yzn, Lattanzi:2017ubx, Chacko:2019nej, Chacko:2020hmh, Escudero:2020ped, FrancoAbellan:2021hdb, Bellomo:2016xhl, Dvali:2016uhn, Lorenz:2018fzb, Dvali:2021uvk, Lorenz:2021alz, Escudero:2022gez, Sen:2023uga, Farzan:2015pca, Alvey:2021xmq, Oldengott:2019lke, Alvey:2021sji, Cuoco:2005qr, Allali:2024anb, Benso:2024qrg}. 

To clarify the robustness of cosmological neutrino-mass determinations, and the complementarity with direct searches, it is key to understand the physical effects of neutrino masses that cosmology is most sensitive to~\cite{Hu:1995em, Audren:2014lsa, Lesgourgues:2014zoa, Lesgourgues:2013sjj}. Is current and upcoming data only sensitive to the background energy in neutrino masses, degenerate with other sources of energy and cosmological unknowns? Or is it sensitive to a characteristic scale-dependent imprint, induced by neutrinos not moving at the speed of light?

In this paper, we explicitly separate background neutrino-mass effects, which capture the evolution of the average neutrino energy density; from perturbations effects, which capture ``kinematic'' scale-dependent effects directly related to the neutrino speed. As a first step, we focus on their impact on CMB anisotropies, leaving other observables for future work~\cite{LSSfuture}. To gain intuition, we study in detail the physical origin of detectable effects. Finally, we carry out an analysis of current CMB data, finding out that it mostly constrains background neutrino-mass effects. In turn, the limit on perturbations effects is significantly weaker than the standard bounds.

Our results provide insight into the physical origin of cosmological neutrino-mass bounds, and serve as a benchmark of extended models that could affect them. As one direct consequence, we find that any change on the background expansion history that is degenerate with the impact of neutrino masses can strongly affect the cosmological neutrino-mass determination. Our definitions of background and perturbations effects are not tied to particular observables (e.g., CMB anisotropies are perturbations, but their evolution is sensitive to the background expansion rate), providing a generic framework to understand the cosmological effects of neutrino masses.

The structure of the manuscript is as follows. In \cref{sec:formalism}, we describe our formalism to separate background and perturbations effects. In \cref{sec:phenomenology}, we examine and illustrate their physical effects on CMB anisotropies at different angular scales. In \cref{sec:results}, we present our statistical analysis together with the cosmological constraints on the key parameters of this study. Finally, we draw our conclusions in \cref{sec:conclusions}. In the Appendices, we provide further explanations and results.

\section{Formalism}
\label{sec:formalism}

After decoupling from the primordial plasma, neutrinos only affect cosmological observables via gravity, i.e., via their stress-energy tensor $T^\mu_\nu$ that enters the Einstein equations. In this Section, we set up the formalism to separate background and perturbations effects. Assuming that neutrinos are non-gravitationally decoupled from photons, baryons, dark matter, and dark energy; we start from the most general parametrization, and compute the relevant quantities for the specific case of massive, non-interacting neutrinos. On top of establishing our framework, this identifies the neutrino-mass effects that cosmology can be sensitive to, enabling a better understanding of the cosmological neutrino-mass determination. 

\subsection{Evolution of the background}

At the background level, the Einstein equations lead to the first Friedmann equation
\begin{equation}\label{eq:Friedmann}
    \left(\frac{a'}{a}\right)^2 = \frac{8\pi G}{3} a^2 (\rho_\gamma+\rho_\mathrm{b}+\rho_\mathrm{cdm}+\rho_\nu+\rho_\Lambda)\, ,
\end{equation}
with primes derivatives with respect to conformal time, $a$ the scale factor, $G$ Newton's constant, and $\rho_i$ the background energy density of the species $i$ (we include photons, baryons, cold dark matter, neutrinos and a cosmological constant). This is the only equation where neutrinos modify the background evolution of the Universe.

The time dependence of the neutrino energy density $\rho_\nu$ is, in turn, governed by the covariant conservation of the neutrino stress-energy tensor
\begin{equation}\label{eq:back_ev}
    \frac{1}{\rho_\nu}\frac{\mathrm{d}\rho_\nu}{\mathrm{d}\ln a} = -3[1+w_\nu(a)]\, ,
\end{equation}
with $w_\nu \equiv P_\nu / \rho_\nu$ the neutrino equation of state and $P_\nu$ the neutrino pressure. The cosmological impact of neutrinos at the background level is fully determined by their initial energy density (usually parametrized by $N_\mathrm{eff}$) and their equation of state, which controls how fast $\rho_\nu$ dilutes.

\begin{figure}[b]
    \centering

    \includegraphics[width=\columnwidth]{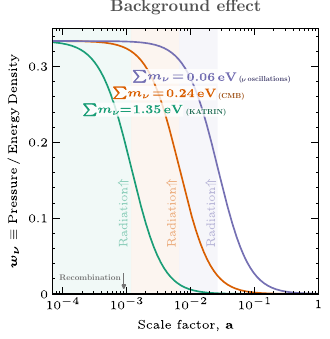}
    \vspace*{-0.55cm}
    \caption{Impact of $\sum m_\nu$ on the neutrino equation of state $w_\nu$, that controls how the neutrino energy density dilutes.  When the neutrino temperature drops below their mass, $w_\nu$ varies from 1/3 (radiation) to 0 (matter). \emph{The only background effect of neutrino masses is to change how fast the neutrino energy dilutes, which modifies the expansion rate of the Universe.}}
    \label{fig:w}
\end{figure}

Differences among particle physics models enter when specifying $w_\nu$. For massive, non-interacting neutrinos that decoupled while being relativistic,
\begin{equation}\label{eq:wnu}
    w_\nu(a, m_\nu) = \frac{1}{3}\frac{\int \mathrm{d}^3p\, \frac{p^2}{\sqrt{p^2+m_\nu^2}}f_0(a p)}{\int \mathrm{d}^3p\, \sqrt{p^2+m_\nu^2}f_0(a p)}\, ,
\end{equation}
with $p$ linear momentum, $m_\nu$ each neutrino mass and
\begin{equation}\label{eq:f0}
    f_0(a p) = \frac{1}{(2\pi)^3} \frac{1}{e^{a p/T_0^\nu}+1}
\end{equation}
a redshifted Fermi-Dirac distribution, with $T_0^\nu \simeq 1.9\,\mathrm{K}$ the current neutrino temperature. \Cref{eq:wnu} encloses the only background effect of neutrino masses.

\Cref{fig:w} shows how, by changing the equation of state, $\sum m_\nu$ determines how fast the neutrino energy density dilutes. We plot the equation of state of non-interacting massive neutrinos for different total neutrino masses. The chosen values are the smallest value allowed by neutrino oscillation experiments, $\sum m_\nu > 0.06\,\mathrm{eV}$~\cite{Esteban:2024eli, deSalas:2020pgw, Capozzi:2018ubv}; the current cosmological limit from Planck CMB data alone, $\sum m_\nu  < 0.24\,\mathrm{eV}$~\cite{Planck:2018vyg}; and the current limit from the KATRIN experiment, $\sum m_\nu  < 1.35\,\mathrm{eV}$~\cite{Katrin:2024tvg}. 

As the figure shows, $\sum m_\nu$ controls at which time $w_\nu$ switches from $1/3$ (radiation, $\rho_\nu \propto a^{-4}$) to $0$ (matter, $\rho_\nu \propto a^{-3}$). The change happens earlier for higher $\sum m_\nu$, leading to a slower dilution of $\rho_\nu$ in \cref{eq:Friedmann} due to the energy in neutrino masses. Hence, the only background effect of increasing neutrino masses, with other parameters fixed, is to increase the expansion rate of the Universe. 

For instance, the background effect of a mass corresponding to current CMB limits (orange line) is that the neutrino energy density dilutes as radiation until $z \sim 10^3$. While in the minimal scenario only neutrino masses control this effect, it is rather indirect, and non-minimal cosmological extensions ---either a different neutrino equation of state or additional background components--- may mimic it~\cite{Chacko:2019nej, Chacko:2020hmh, Escudero:2020ped, FrancoAbellan:2021hdb, Bellomo:2016xhl, Dvali:2016uhn, Lorenz:2018fzb, Dvali:2021uvk, Lorenz:2021alz, Escudero:2022gez, Esteban:2021ozz, Esteban:2022rjk, Sen:2023uga, Farzan:2015pca, Alvey:2021xmq, Oldengott:2019lke, Alvey:2021sji, Cuoco:2005qr, Allali:2024anb, Benso:2024qrg}. 

Below, we separate this effect from perturbations effects. To do so, we parametrize $w_\nu$ as in the massive, non-interacting case $w_\nu = w_\nu(a, \sum m_\nu^\mathrm{Backg.})$; with $\sum m_\nu^\mathrm{Backg.}$ a parameter that we name \emph{background neutrino mass}.

\subsection{Evolution of perturbations}
\label{sec:pert}

\begin{figure*}[ht]
    \centering
    \includegraphics[width=\textwidth]{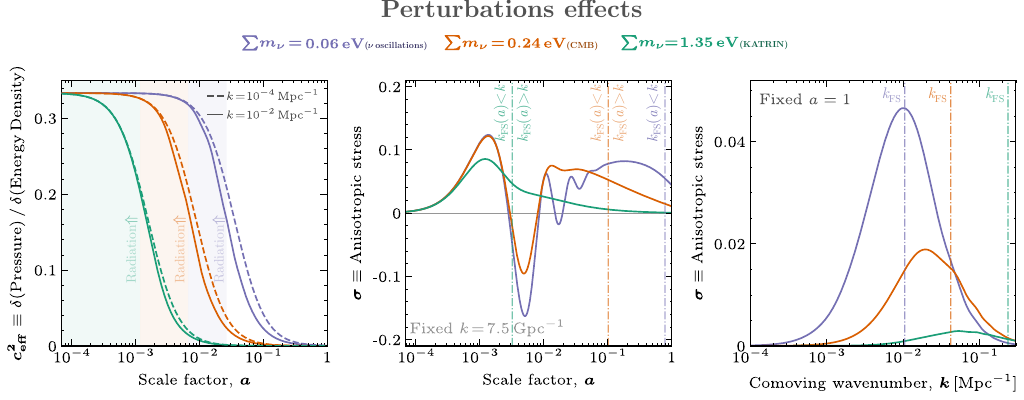} 
    \caption{Impact of $\sum m_\nu$ on the comoving neutrino squared-sound-speed $c_{\mathrm{eff}}^2$ [left] and anisotropic stress $\sigma$ [center, for fixed scale and varying time; right, for fixed time and varying scale]. These parameters control all perturbations effects of neutrinos (see text). \emph{$c_\mathrm{eff}^2$ is essentially scale-independent and falls when the neutrino temperature drops below their mass. In turn, $\sigma$ gets suppressed above a characteristic scale, the neutrino free-streaming length, that depends on the neutrino mass.}}
    \label{fig:pert}
\end{figure*}

At the perturbations level, for scalar perturbations the neutrino stress-energy tensor $T^\mu_\nu$ can be written in terms of the energy density perturbation $\delta \rho$, the pressure perturbation $\delta P$, the velocity divergence $\theta$, and the anisotropic stress $\sigma$. In Fourier space,~\cite{Ma:1995ey}
\begin{align}
    \delta \rho & \equiv - T^0_0 - \rho_\nu \, , \\
    \delta P & \equiv \frac{1}{3}T^i_i - P_\nu \, , \\
    \theta & \equiv \frac{i k^{i} T^0_i}{\rho_\nu + P_\nu} \, , \\
    \sigma & \equiv - \frac{\hat{k}_i \hat{k}_j [T^i_j - (P_\nu + \delta P) \delta^i_j]}{\rho_\nu + P_\nu} \, ,
\end{align}
with $k$ the comoving wavenumber. Physically, $\theta$ represents the divergence of the bulk velocity of energy perturbations; and $\delta P$ and $\sigma$ the isotropic and anisotropic components of the linear momentum flux, respectively. 

These parameters source the perturbed Einstein equations~\cite{Ma:1995ey}. They evolve following energy-momentum conservation, which in the conformal Newtonian gauge reads
\begin{align}
    \delta' &= -(1+w_\nu)\theta +3(1+w_\nu)\phi' - 3 \frac{a'}{a} \left(c_s^2 - w_\nu\right) \delta \, , \label{eq:delta_nu} \\
    \theta' &= -\frac{a'}{a}(1 - 3 c_\mathrm{ad}^2) \theta + \frac{c_s^2}{1+w_\nu}k^2\delta - k^2 \sigma + k^2 \psi\, ,\label{eq:theta_nu}
\end{align}
where $\delta \equiv \delta \rho / \rho_\nu$, $\phi$ and $\psi$ are the gravitational potentials in the conformal Newtonian gauge, $c_s^2 \equiv \delta P / \delta \rho$ is the so-called squared sound-speed, and 
\begin{equation} \label{eq:ad}
    c_\mathrm{ad}^2 \equiv \frac{P_\nu'}{\rho_\nu'} = w_\nu - \frac{w_\nu'}{3 \frac{a'}{a}(1+w_\nu)} \, ,
\end{equation} 
is the so-called adiabatic squared sound-speed. Physically, the first three terms in \cref{eq:delta_nu} correspond to energy dilution due to bulk motions, gravitational redshift, and the expansion of the Universe; respectively. In \cref{eq:theta_nu}, the first term corresponds to drag due to the expansion of the Universe; the second and third terms to isotropic and anisotropic momentum flow, respectively; and the fourth term to gravitational forces. 

To solve these equations, $c_s^2$ and $\sigma$ (together with the background quantity $w_\nu$) have to be provided. An indeterminacy arises because $c_s^2$ is gauge-dependent, that is, its value depends on the coordinate system used to separate background from perturbations. Physically, by performing a coordinate transformation one can transfer pressure and energy density from perturbations to the background and vice versa. A related issue is that, for adiabatic perturbations that behave as background as $k \rightarrow 0$, the evolution equations for small $k$ should only depend on background quantities. This may enforce a consistency relation between $c_s^2$, $\sigma$, and $w_\nu$. These issues can be overcome if, instead of $c_s^2$, the equations are written in terms of the so-called effective sound-speed~\cite{Hu:2004kh, Garriga:1999vw, Hu:1998kj}
\begin{equation}\label{eq:ceff_nu}
    c_\mathrm{eff}^2 \equiv \frac{k^2 c_s^2\delta + 3\frac{a'}{a}(1+w_\nu)c_\mathrm{ad}^2\theta}{k^2\delta+3\frac{a'}{a}(1+w_\nu)\theta} \, .
\end{equation}
As we show in~\cref{sec:bkg-consistency}, $c_\mathrm{eff}^2$ is gauge-invariant and, when expressing the evolution equations in terms of it, adiabatic perturbations behave as background as $k \rightarrow 0$ \emph{regardless of the values of $c_\mathrm{eff}^2$, $\sigma$, and $w_\nu$}. Physically, $c_\mathrm{eff}$ is the sound speed in a frame comoving with neutrinos.

The equations above always hold --- assuming, as mentioned above, that neutrinos are decoupled from standard species. Differences among particle physics models (including neutrino-mass effects) enter when specifying $c_\mathrm{eff}^2$ and $\sigma$. For massive, non-interacting neutrinos; they can be computed from the perturbed distribution function
\begin{align}\label{eq:f0pert}
    f(\vec{k},\vec{p},\eta) = f_0(ap)\left[1+\Psi(\vec{k},\vec{p},\eta)\right]\, ,
\end{align}
with $\eta$ conformal time. The explicit expression for the stress-energy tensor leads to~\cite{Ma:1995ey}
\begin{align}
    \delta &= \frac{1}{\rho_\nu} \int \mathrm{d}^3p\,\sqrt{p^2+m_\nu^2}f_0 \Psi\, ,\label{eq:deltatower}\\
    \theta &= \frac{1}{\rho_\nu + P_\nu} \int \mathrm{d}^3p\, (i \vec{k} \cdot \vec{p}) f_0 \Psi\, ,\label{eq:thetatower}\\
    \delta P &= \frac{1}{3}\int \mathrm{d}^3p\,p v f_0\Psi\, ,\label{eq:deltaPtower}\\
    \sigma &= \frac{1}{\rho_\nu + P_\nu} \int \mathrm{d}^3p\,p v \left[\frac{1}{3}-(\hat{k}\cdot \hat{p})^2\right]f_0\Psi\label{eq:sheartower}\,, 
\end{align}
where $v \equiv p/\sqrt{p^2+m_\nu^2}$ is the neutrino velocity. These expressions explicitly show the physical meaning of $\theta$, $\delta P$, and $\sigma$ for massive, non-interacting neutrinos. The evolution of $\Psi$ follows the perturbed Boltzmann equation~\cite{Ma:1995ey}.

\Cref{fig:pert} shows that the anisotropic stress contains the leading ``kinematic'' effects of $\sum m_\nu$. We plot the time- and scale-dependence of $c_\mathrm{eff}^2$ and $\sigma$ (normalized to an initial comoving curvature perturbation $\mathcal{R} = 1$) of non-interacting massive neutrinos, for the same total neutrino masses as in \cref{fig:w}. We fix other cosmological parameters to the best fit of the Planck 2018 CMB analysis~\cite{Planck:2018vyg}.

The scale dependence can be understood in terms of the neutrino free-streaming wavenumber~\cite{Lesgourgues:2013sjj}
\begin{equation}\label{eq:k_FS}
    k_\mathrm{FS}(a) \simeq 0.776 \frac{a^2 H(a)}{H_0}\left(\frac{m_\nu}{1\,\mathrm{eV}}\right) h~\mathrm{Mpc^{-1}}\, ,
\end{equation}
with $H$ the Hubble parameter and $h$ the reduced Hubble constant. Physically, if $k < k_\mathrm{FS}$, perturbations tend to collapse gravitationally; whereas if $k > k_\mathrm{FS}$, velocity dispersion inhibits gravitational collapse~\cite{Lesgourgues:2013sjj}. Thus, $k_\mathrm{FS}$ directly encodes the ``kinematic'' impact of $\sum m_\nu \neq 0$, i.e., that neutrinos do not move at the speed of light.

As the left panel of \cref{fig:pert} shows, the time dependence of $c_\mathrm{eff}^2$ resembles that of the equation of state, falling from $1/3$ to $0$ when neutrinos become non-relativistic. The scale dependence introduced by $k_\mathrm{FS}$ is subleading (see Ref.~\cite{Nascimento:2023psl} for a discussion). Physically, super-horizon adiabatic perturbations behave as background, $c_\mathrm{eff}^2 \rightarrow c_\mathrm{ad}^2$ at all scales, and scale dependence only appears after subleading sub-horizon evolution~\cite{Nascimento:2023psl, Lesgourgues:2011rh}.

The center and right panels of \cref{fig:pert} show that the evolution of $\sigma$ is much more scale-dependent, with a characteristic feature at $k = k_\mathrm{FS}$. As the center panel shows, at early times $\sigma$ oscillates. These oscillations get steadily damped with time, as neutrino free-streaming steadily suppresses perturbations. However, when the mode becomes larger than the free-streaming scale (which shrinks with time as neutrinos become non-relativistic), i.e., when $k < k_\mathrm{FS}$; neutrinos cluster instead of free-streaming, the momentum flux diminishes, and $\sigma$ decays much faster. The right panel also shows the two distinct behaviors as a function of scale at fixed time. At scales below the free-streaming scale, $k > k_\mathrm{FS}$, $\sigma$ is larger at large scales, which had less time to evolve and are less damped by free-streaming. On the contrary, at scales above the free-streaming scale, $k < k_\mathrm{FS}$, $\sigma$ is smaller at large scales, where neutrino clustering reduces the momentum flux. These distinct behaviors are a direct consequence of neutrinos not moving at the speed of light.

To separate the background and perturbations effects of neutrino masses, below we parametrize $c_\mathrm{eff}^2$ and $\sigma$ as in the massive, non-interacting case; computing them for a value of $\sum m_\nu$ that we name \emph{perturbations neutrino mass}, $\sum m_\nu^\mathrm{Pert.}$ (see \cref{sec:nu_evo} for details). Given $c_\mathrm{eff}^2$ and $\sigma$, we solve \cref{eq:delta_nu,eq:theta_nu}, that only assume stress-energy conservation and are therefore exact. We then use the result to source the Einstein equations.

\section{Phenomenology}
\label{sec:phenomenology}

The formalism described above generically separates neutrino background effects, parametrized by $w_\nu$; and perturbations effects, parametrized by $c_\mathrm{eff}^2$ and $\sigma$. In this Section, we use this formalism to separate the background and perturbations effects of neutrino masses. We discuss the physical effects on CMB anisotropies, their origin, and how they can be observationally identified.

\subsection{Implementation}
\label{sec:implementation}

To implement background effects, we solve \cref{eq:back_ev} for an equation of state of massive, non-interacting neutrinos with total mass $\sum m_\nu^\mathrm{Backg.}$,
\begin{equation}
    \rho_\nu(a) = \rho_\nu(a_1) \exp\left[ -3 \int_{a_1}^{a} \frac{1 + w_\nu(a, \sum m_\nu^\mathrm{Backg.})}{a} \, \mathrm{d}a\right] \, ,
\end{equation}
where $a_1$ is an initial scale factor. For simplicity, we assume the standard initial energy density for three light neutrino species,~\cite{Lesgourgues:2013sjj}
\begin{equation}
    \rho_\nu(a_1) = \frac{7 \pi^2}{40 a_1^4} (T_0^\nu)^4 \, ,
\end{equation}
with the initial condition evaluated when neutrinos are ultrarelativistic, i.e., $T_0^\nu/a_1 \gg \sum m_\nu^\mathrm{Backg.}$. The hereby computed neutrino energy density affects the expansion of the Universe via the Friedmann equation, \cref{eq:Friedmann}.

To implement perturbations effects, we solve \cref{eq:delta_nu,eq:theta_nu} for values of $c_\mathrm{eff}^2$ and $\sigma$ corresponding to neutrinos with total mass $\sum m_\nu^\mathrm{Pert.}$, together with the perturbations equations of other species. We set initial conditions corresponding to adiabatic perturbations of an ultrarelativistic relic~\cite{Ma:1995ey}. We provide the technical details of our computation in~\cref{sec:nu_evo}. The hereby computed neutrino energy density and pressure perturbations $\delta \rho_\nu$ and $\delta P_\nu$, velocity divergence $\theta$, and anisotropic stress $\sigma$ affect the evolution of the Universe via the perturbed Einstein equations. We implement our modified cosmological evolution in the public code \texttt{CLASS}~\cite{Lesgourgues:2011re, Blas:2011rf, Lesgourgues:2011rg, Lesgourgues:2011rh}. 

\subsection{Physical effects on the CMB}
\label{sec:CMB}

\begin{figure}[b]
    \centering

    \includegraphics[width=\columnwidth]{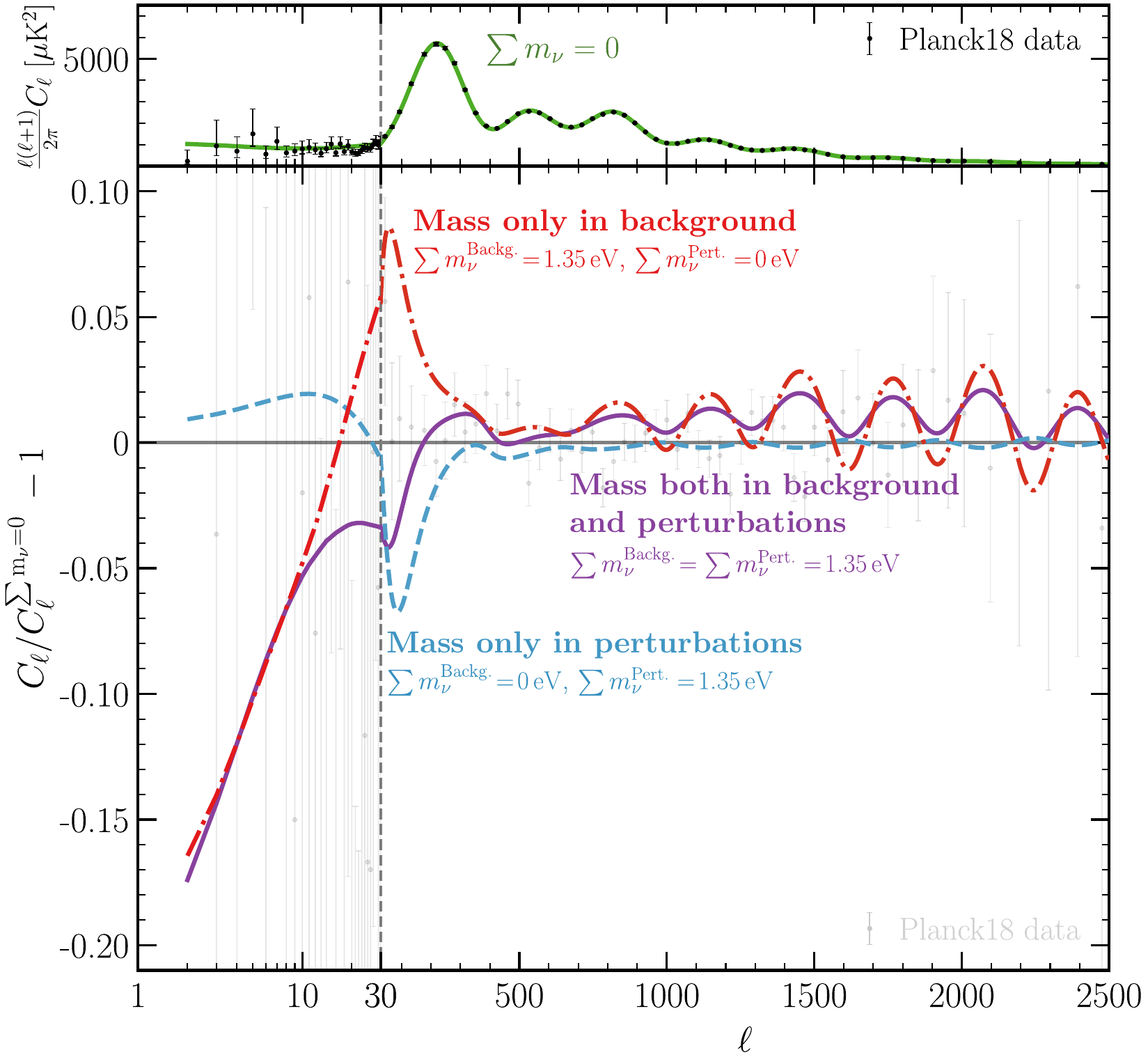}
    
    \caption{Background- and perturbations-induced impact of $\sum m_\nu$ on CMB anisotropies. As detailed in the main text; at low and high $\ell$, the background affects the LISW effect, Silk damping, and lensing; at intermediate $\ell$, both background and perturbations directly couple to photon-baryon oscillations via gravity. \emph{Most effects are background-induced. Perturbations-induced effects are mainly relevant at intermediate $\ell$, where their effect is opposite to that of the background.}}

    \label{fig:cmb1}
\end{figure}

\Cref{fig:cmb1} shows that $\sum m_\nu^\mathrm{Backg.}$ and $\sum m_\nu^\mathrm{Pert.}$ leave distinct signatures on CMB anisotropies at different scales. To better represent the observable effects, in the figure we fix the well-measured cosmological parameters $\{100\theta_s,\,\omega_b,\,\omega_{\mathrm{cdm}},\,A_s,\,n_s,\,\tau_{\mathrm{reio}}\}$ to the best fit of the Planck 2018 CMB analysis~\cite{Planck:2018vyg}. Below, we explore in detail the physical origin of the different effects. In~\cref{sec:app_CMB}, we provide explicit checks and figures that further illustrate our results.

Firstly, low multipoles ($\ell \lesssim 10$) correspond to modes that enter the horizon at very late times, when neutrinos constitute a subleading component of the energy density of the Universe. The main neutrino-mass effects are thus induced when the modes are larger than the horizon. Since, for adiabatic perturbations, the evolution of super-horizon modes depends only on background quantities, the effect of $\sum m_\nu^\mathrm{Pert.}$ at these scales is negligible. The main effect of $\sum m_\nu^\mathrm{Backg.}$ is indirect: an increased $\sum m_\nu^\mathrm{Backg.}$ slows down the dilution of the neutrino energy density, increasing $H(z)$ and, in principle, modifying the observed angular scale of CMB peaks~\cite{Lesgourgues:2013sjj}
\begin{equation} \label{eq:theta_s}
\theta_s = \left.\int_{z_\mathrm{rec}}^\infty \frac{c_{s,\gamma}(z)\, \mathrm{d}z}{H(z)} \right/ \int_0^{z_\mathrm{rec}} \frac{\mathrm{d}z}{H(z)}  \, ,
\end{equation}
with $c_{s,\gamma}(z)$ the sound speed of the baryon-photon fluid and $z_\mathrm{rec}$ the redshift of recombination. Since $\theta_s$ is very-well measured, changes in it are compensated by modifying $H_0$, which modifies the cosmological constant $\Lambda$. This changes the $\Lambda$-induced late-time boosting of large-scale anisotropies through the late integrated Sachs-Wolfe (LISW) effect~\cite{Sachs:1967er}. In addition, the amplitude of super-horizon CMB perturbations depends on the expansion rate of the Universe around recombination (see~\cref{sec:app_CMB}), which depends on $\sum m_\nu^\mathrm{Backg.}$. This introduces a subleading depletion of low-$\ell$ CMB anisotropies.

Secondly, high multipoles ($\ell \gtrsim 500$) correspond to modes that enter the horizon much before recombination. There are two main ways in which neutrino masses affect these modes, both of which are mainly sensitive to background effects. 

On the one hand, these modes are damped below a characteristic angular scale $\theta_D$ due to the finite mean free path of photons, where~\cite{Lesgourgues:2013sjj}
\begin{equation}\label{eq:theta-damping}
    \theta_D \sim \left.\sqrt{\int_{z_\mathrm{rec}}^\infty \frac{1+z}{n_e(z) \sigma_T} \frac{\mathrm{d}z}{H(z)}} \right/ \int_0^\mathrm{z_\mathrm{rec}} \frac{\mathrm{d}z}{H(z)}  \, ,
\end{equation}
with $n_e$ the electron density and $\sigma_T$ the Thomson scattering cross section. An increased $\sum m_\nu^\mathrm{Backg.}$ slows down the dilution of the neutrino energy density, increasing $H(z)$ both in the numerator and denominator. The change in $H_0$ to keep $\theta_s$ fixed, discussed above, further affects the denominator. Overall, $\theta_\mathrm{D}$ gets reduced, which is visible as an excess at high $\ell$ in \cref{fig:cmb1}. As the damping scale only depends on background quantities, this effect is insensitive to $\sum m_\nu^\mathrm{Pert.}$ (see \cref{sec:app_CMB}). 

On the other hand, high multipoles are affected by weak gravitational lensing, i.e., by the random gravitational deflection of CMB photons due to the large-scale structure of the Universe. Lensing smooths out the power spectrum and transfers power from low multipoles to high multipoles~\cite{Dodelson:2020abc}. Neutrino masses affect CMB lensing via both background and perturbations effects (see \cref{sec:app_CMB} for details). $\sum m_\nu^\mathrm{Backg.}$ reduces CMB lensing by accelerating the expansion of the Universe, which suppresses structure formation. This is visible in \cref{fig:cmb1} as wiggles that are in phase with the CMB power spectrum. In turn, $\sum m_\nu^\mathrm{Pert.}$ enhances neutrino clustering and structure growth, which enhances CMB lensing in a scale-dependent way. This effect is relevant if $\sum m_\nu^\mathrm{Backg.} \neq 0$, because otherwise the neutrino energy density is too diluted at late times and neutrino perturbations effects are negligible. Overall, the main impact of $\sum m_\nu^\mathrm{Pert.}$ on CMB lensing is to partially reduce the effect of $\sum m_\nu^\mathrm{Backg.}$ as can be seen in \cref{fig:cmb1}. In our data analysis below, this leads to a partial degeneracy among both parameters.

Finally, intermediate multipoles ($10\lesssim \ell \lesssim 500$) correspond to modes that enter the horizon around recombination. Anisotropies at these scales are largely influenced by the gravitational potentials around recombination, that act as a driving term for photon-baryon acoustic oscillations. More precisely, decaying gravitational potentials increase the amplitude of oscillations~\cite{Hu:1998kj}. $\sum m_\nu^\mathrm{Backg.}$ boosts this decay by increasing the expansion rate of the Universe, thus boosting the amplitude of anisotropies as can be seen in \cref{fig:cmb1}. In turn, $\sum m_\nu^\mathrm{Pert.}$ enhances neutrino clustering and structure growth in a scale-dependent way (see \cref{sec:app_CMB}), slowing down the decay of the gravitational potentials and decreasing the amplitude of anisotropies. On top of these effects, decaying gravitational potentials further increase the amplitude of CMB anisotropies through the integrated Sachs-Wolfe effect, boosting the aforementioned effects (see \cref{sec:app_CMB}). The fact that the effects of $\sum m_\nu^\mathrm{Backg.}$ and $\sum m_\nu^\mathrm{Pert.}$ are opposite at intermediate $\ell$ leads to a partial degeneracy among both parameters in our data analysis.

We conclude that, while background effects are present at all multipoles, $\sum m_\nu^\mathrm{Pert.}$ on its own is mainly relevant at intermediate $\ell$, where anisotropies are suppressed due to the direct gravitational impact of neutrino perturbations. Even if this effect can be partially hidden by a non-zero $\sum m_\nu^\mathrm{Backg.}$, unique background effects at high $\ell$ allow disentangling both. In the next section, we carry out a data analysis to quantify the effects allowed by current CMB data. The higher statistics at large $\ell$ forecasts a stronger limit on background effects.

\section{Data analysis and results}
\label{sec:results}

In the Sections above, we have discussed how the different neutrino masses that we introduce affect the cosmological evolution. In short, $\sum m_\nu^\mathrm{Backg.}$ encodes the equation of state, i.e., how fast the neutrino energy density dilutes; while $\sum m_\nu^\mathrm{Pert.}$ contains a more direct ``kinematic'' effect related to the free-streaming nature of neutrinos. In this Section, we carry out an analysis of CMB data to quantify the allowed values of $\sum m_\nu^\mathrm{Backg.}$ and $\sum m_\nu^\mathrm{Pert.}$.

We analyze the Planck 2018 temperature, polarization, and lensing power spectra (\texttt{TT, TE, EE+lowE+lensing} in Ref.~\cite{Planck:2018vyg}). To do so, we modify the public code \texttt{CLASS}~\cite{Lesgourgues:2011re, Blas:2011rf, Lesgourgues:2011rg, Lesgourgues:2011rh}, as mentioned above and detailed in \cref{sec:nu_evo}, to solve the evolution of cosmological perturbations; and we explore the parameter space with the public Markov Chain Monte Carlo code \texttt{COBAYA}~\cite{Torrado:2020dgo, 2019ascl.soft10019T}. In \cref{sec:app_full_analysis}, we summarize our priors and convergence criteria, and we provide the full results of our analysis.

\begin{figure}[t]
    \centering

    \includegraphics[width=\columnwidth]{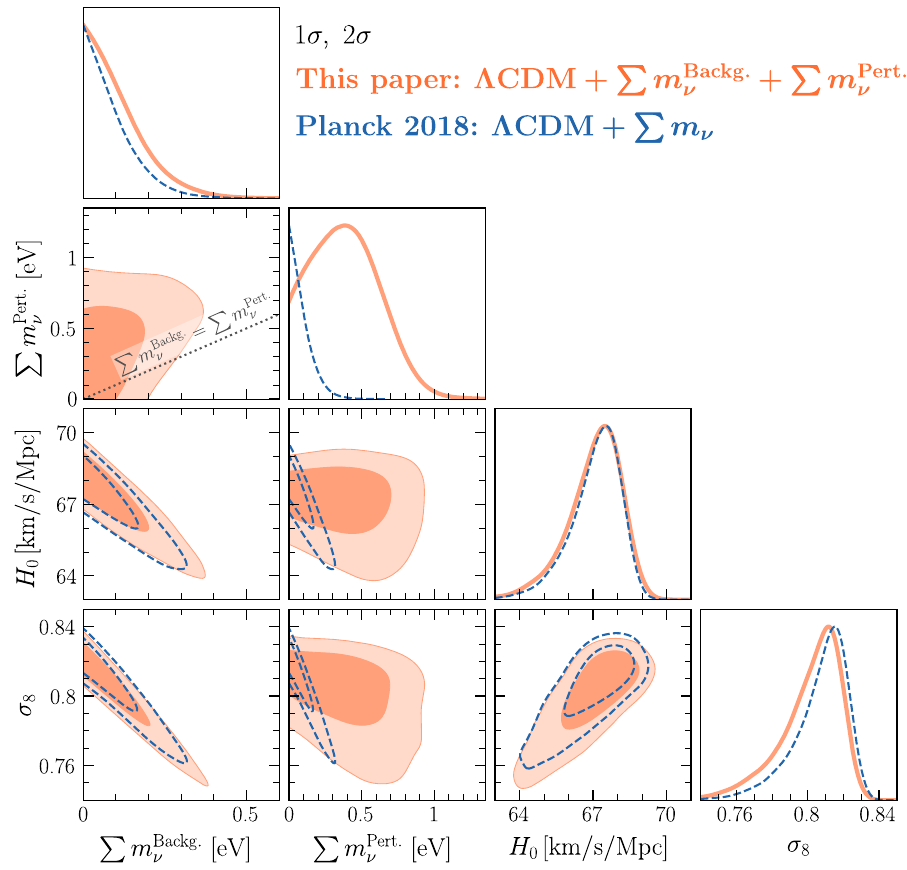}
    
    \caption{CMB limits on the separate neutrino-mass effects due to background and perturbations. The limit on $\sum m_\nu^\mathrm{Pert.}$ gets strongly relaxed compared to that on $\sum m_\nu$. \emph{The CMB $\sum m_\nu$ limit  is mostly a limit on background neutrino-mass effects.}}

    \label{fig:triangle}
\end{figure}

\cref{fig:triangle} shows that splitting neutrino mass effects among background and perturbations strongly increases the allowed perturbations effects. We show in solid the 1D posterior probabilities and 2D credible regions of our analysis for $\sum m_\nu^\mathrm{Backg.}$, $\sum m_\nu^\mathrm{Pert.}$, the Hubble parameter $H_0$, and the amplitude parameter $\sigma_8$. Dashed lines correspond to the standard scenario, i.e., $\sum m_\nu^\mathrm{Backg.} = \sum m_\nu^\mathrm{Pert.}$. 

The $95\%$ CL limits on the total neutrino mass read
\begin{align}
    &\sum m_\nu< 0.24~\mathrm{eV} & \mathrm{({\color{Planck18}Planck~2018}~\text{\cite{Planck:2018vyg}})}\,, \nonumber\\
    &\sum m_\nu^\mathrm{Backg.} < 0.29~\mathrm{eV} & \mathrm{({\color{Thispaper}This~paper})}\,, \nonumber \\
    &\sum m_\nu^\mathrm{Pert.} < 0.79~\mathrm{eV} & \mathrm{({\color{Thispaper}This~paper})}\,. \nonumber 
\end{align}
That is, the standard limit on $\sum m_\nu$ is mostly a limit on $\sum m_\nu^\mathrm{Backg.}$ (the limit on the latter is slightly weaker due to a degeneracy with $\sum m_\nu^\mathrm{Pert.}$ that we discuss below). As the figure shows, the posterior probabilities of $\sum m_\nu$, in the standard analysis; and $\sum m_\nu^\mathrm{Backg.}$, in our analysis; almost match. The correspondence among $\sum m_\nu$ and $\sum m_\nu^\mathrm{Pert.}$, however, is null. 

In other words, CMB data tightly constrains the neutrino equation of state; but, compared to the standard scenario, the limit on ``kinematic'' effects of neutrino masses is relaxed by about a factor of 3.

The background neutrino mass is correlated with $H_0$ and $\sigma_8$. These correlations are also present in the standard scenario, and they are due to the neutrino contribution to the total energy density of the Universe. Larger $\rho_\nu$ increases the expansion rate, which suppresses structure formation (i.e., $\sigma_8$); and, as described in the previous Section, modifies the angular scale of the CMB peaks that is degenerate with $H_0$. These are both background effects, so $\sum m_\nu^\mathrm{Pert.}$ is not strongly correlated with $H_0$ and $\sigma_8$.

\begin{figure}[t]
    \centering

    \includegraphics[width=\columnwidth]{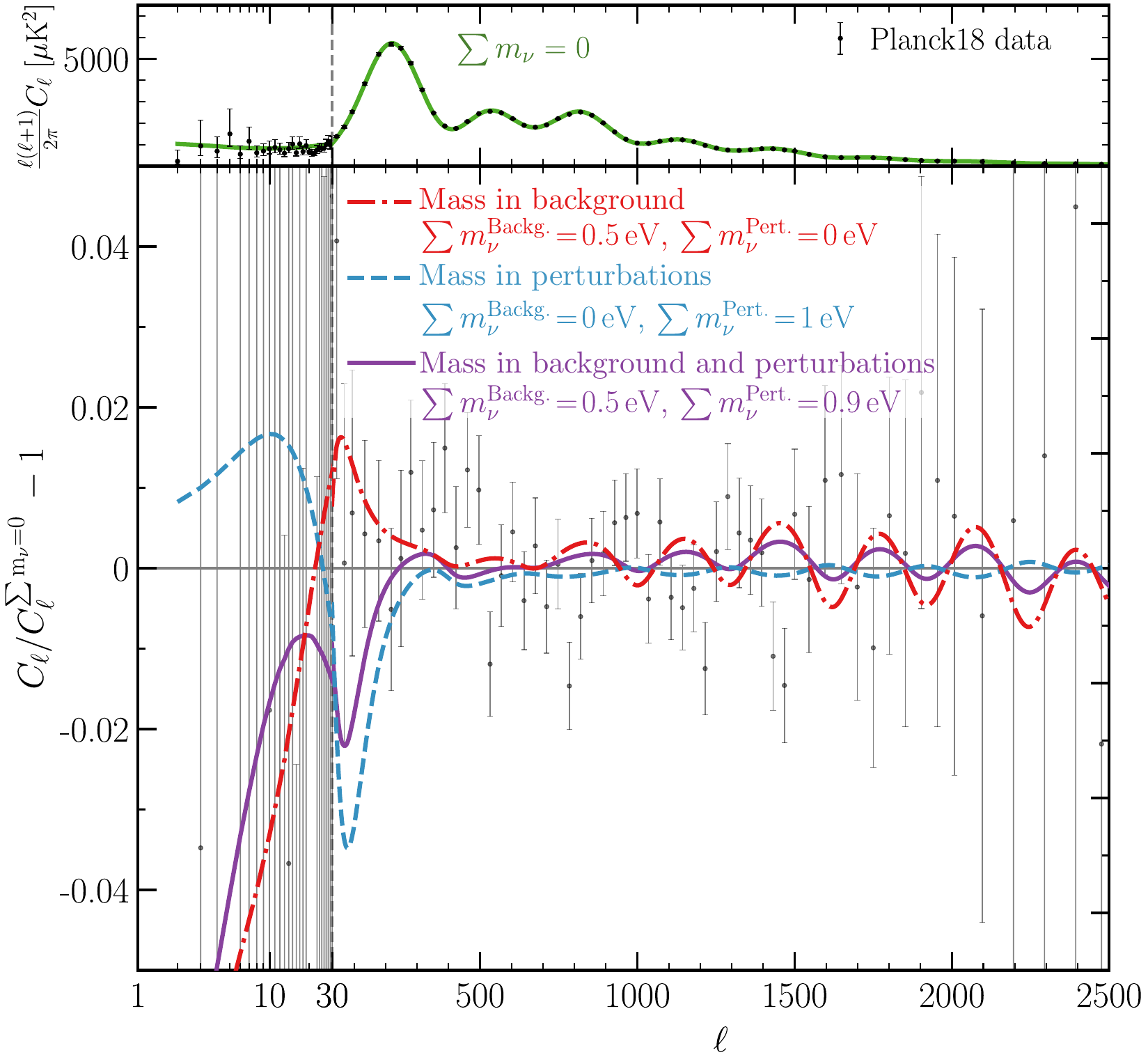}
    
    \caption{Impact on CMB anisotropies of parameters excluded by our analysis. Background effects are most excluded by high-$\ell$ data (CMB lensing), and perturbations effects by intermediate-$\ell$ data (direct coupling via gravity). \emph{Opposed physical effects induce a degeneracy between $\sum m_\nu^\mathrm{Pert.}$ and $\sum m_\nu^\mathrm{Backg.}$ (see text).}}
    \label{fig:cmb2}
\end{figure}

\Cref{fig:cmb2} shows that intermediate-$\ell$ data constrains both perturbations and background mass effects, whereas high-$\ell$ data mainly constrains background-mass effects. In both cases, perturbations-mass effects can partially compensate background mass effects; we discuss this below. As in \cref{fig:cmb1}, we fix $\{100\theta_s,\,\omega_b,\,\omega_{\mathrm{cdm}},\,A_s,\,n_s,\,\tau_{\mathrm{reio}}\}$ to their Planck 2018 best-fit values~\cite{Planck:2018vyg}. The blue and purple lines are close to the $2\sigma$ allowed region in our analysis, whereas the red line is more strongly excluded.

As discussed in the previous Section, the sensitivity at intermediate multipoles is due to the direct impact of neutrino masses on the gravitational potentials. The high-$\ell$ sensitivity to background mass effects can be traced back to the impact on the damping tail and CMB lensing. This is particularly relevant in light of the lensing anomaly: Planck 2018 data prefers more CMB lensing than what is present in the standard $\Lambda$CDM scenario~\cite{Planck:2018vyg,Motloch:2018pjy,Motloch:2019gux,Efstathiou:2019mdh,SPT-3G:2024atg} (see, however, Refs.~\cite{Tristram:2023haj, Rosenberg:2022sdy}). Since, as described in the previous Section, $\sum m_\nu^\mathrm{Backg.}$ \emph{reduces} CMB lensing, the anomaly enhances the high-$\ell$ constraints, and the resulting limit is somewhat stronger than what would be expected. This is visible in the bottom panel of \cref{fig:cmb2} as wiggles in the data that are out of phase with the effect of $\sum m_\nu^\mathrm{Backg.}$.

An interesting feature of \cref{fig:triangle}, also visible in \cref{fig:cmb2}, is that when splitting neutrino masses into background and perturbations, $\sum m_\nu^\mathrm{Backg.}$ and $\sum m_\nu^\mathrm{Pert.}$ are partly degenerate. When $\sum m_\nu^\mathrm{Pert.}$ increases, the limit over $\sum m_\nu^\mathrm{Backg.}$ can relax by almost a factor of two. This is driven by opposite effects of background and perturbations: $\sum m_\nu^\mathrm{Backg.}$ suppresses structure formation due to the increased expansion rate of the Universe, whereas $\sum m_\nu^\mathrm{Pert.}$ enhances structure formation due to neutrinos clustering as matter. As described in the previous Section, this introduces opposite effects on the amplitude of CMB anisotropies, at intermediate $\ell$; and on CMB lensing, at high $\ell$. The degeneracy is not perfect, partly because increasing $\sum m_\nu^\mathrm{Backg.}$ also affects the CMB damping tail that is insensitive to $\sum m_\nu^\mathrm{Pert.}$.

Overall, our results explicitly show that, while the CMB is sensitive both to background and perturbations effects, the limit on $\sum m_\nu$ is roughly dominated by background effects, i.e., by how the neutrino energy density evolves with time. Once we separate neutrino-mass effects into background and perturbations (the latter containing more direct ``kinematic'' signatures related to neutrino free-streaming), the limit on perturbations effects gets largely relaxed. This could serve as an insight for theories that evade cosmological neutrino mass bounds, as our results show that such theories should have a radiation-like dilution of the neutrino energy density until recombination.

\section{Conclusions and Outlook}
\label{sec:conclusions}

The absolute neutrino mass scale remains unknown. Cosmological surveys are close to either measuring it or excluding the lower limit set by oscillations. In this paper, we explore the origin of the cosmological neutrino-mass bound from CMB data. Our results show that the driving constraint arises from the contribution of the energy in neutrino masses to the expansion of the Universe. 

We focus on the macroscopic quantities that capture the cosmological impact of neutrinos. This enables us to explicitly discriminate, for the first time, between background and perturbations effects of neutrino masses. The separation can be understood in terms of standard ``fluid'' variables: the equation of state $w_\nu$, which governs how fast the background neutrino energy density dilutes; the sound speed in the frame comoving with neutrinos $c_\mathrm{eff}^2$, which captures isotropic neutrino momentum flow; and the anisotropic stress $\sigma$, which captures anisotropic neutrino momentum flow. $w_\nu$ encloses all background effects, directly impacting the expansion rate of the Universe---which increases as $\sum m_\nu$ increases due to the energy in neutrino masses. In turn, $c_\mathrm{eff}^2$ and $\sigma$ enclose the perturbations effects. $c_\mathrm{eff}^2$ is almost scale-independent, and $\sigma$ contains the main ``kinematic'' impact of neutrino masses, related to the free-streaming scale set by neutrinos not moving at the speed of light.

Since our goal is to disentangle among background and perturbations effects, we explore two types of neutrino masses: the one that governs $w_\nu$, that we name \emph{background neutrino mass} $\sum m_\nu^\mathrm{Backg.}$; and the one that governs $c_\mathrm{eff}^2$ and $\sigma$, that we name \emph{perturbations neutrino mass} $\sum m_\nu^\mathrm{Pert.}$. Although these parameters are phenomenological, they encode the distinct physical implications of neutrino masses in cosmology. Hence, they serve as a benchmark to understand the effects that cosmology is most sensitive to, and to shed light on potential degeneracies of extended models with $\sum m_\nu$.

The effects on the CMB temperature anisotropies can be split into different multipole regions (see \cref{sec:CMB}). The low-$\ell$ region has a minor impact on the constraints due to cosmic variance. At high $\ell$, $\sum m_\nu^\mathrm{Backg.}$ modifies the CMB damping tail, and both background and perturbations effects impact CMB lensing in opposite directions. While a larger $\sum m_\nu^\mathrm{Backg.}$ suppresses structure formation (and hence CMB lensing) by increasing the expansion rate of the Universe, a larger $\sum m_\nu^\mathrm{Pert.}$ enhances structure formation above the neutrino free-streaming scale. At intermediate $\ell$, neutrino perturbations directly affect photon-baryon oscillations via gravity, where again the effects of $\sum m_\nu^\mathrm{Backg.}$ and $\sum m_\nu^\mathrm{Pert.}$ are opposite. As a consequence, there is a slight degeneracy among $\sum m_\nu^\mathrm{Backg.}$ and $\sum m_\nu^\mathrm{Pert.}$. Overall, high-$\ell$ data is mostly sensitive to background effects, whereas intermediate-$\ell$ data determines both background and perturbations effects.

We then carry out an analysis of Planck 2018 CMB data to shed light on the effects that observations constrain. We conclude that the Planck 2018 neutrino-mass bound is a bound on the background effects, i.e., on the evolution of the neutrino energy density. This provides a rule-of-thumb to understand if CMB data excludes a model with new physics in the neutrino sector: if its equation of state significantly deviates from $w_\nu=1/3$ around recombination, the model is probably excluded.

The perturbations limit on ``kinematic'' effects of neutrino masses is consequently relaxed, $\sum m_\nu^\mathrm{Pert.} < 0.8\,\mathrm{eV}$. The limit is still competitive---models that dramatically affect free-streaming properties of neutrinos are still excluded---, and it is similar to the projected reach of KATRIN~\cite{Katrin:2024tvg}; yet in the standard scenario such high neutrino masses are excluded within $\sim 7\sigma$. This result underscores the complementarity among laboratory and cosmological determinations of the neutrino mass.

In this first paper, we have focused on the consequences for the CMB of ``energy-dilution'' versus ``kinematic'' effects of neutrino masses. Since current and near-future observations of the matter power spectrum (particularly Baryon Acoustic Oscillation, BAO, measurements) have a strong impact on neutrino-mass determinations~\cite{DESI:2024mwx, DESI:2024hhd}, we will explore them in detail in incoming work~\cite{LSSfuture}. For BAOs, which measure quantities that can be expressed in terms of background neutrino properties~\cite{Esteban:2021ozz}, we foresee a stronger impact on $\sum m_\nu^\mathrm{Backg.}$ than on $\sum m_\nu^\mathrm{Pert.}$.

Our separation among background and perturbation neutrino-mass effects opens many research avenues. The observed CMB-lensing excess drives the strong limit on $\sum m_\nu^\mathrm{Backg.}$ and the best-fit for nonzero $\sum m_\nu^\mathrm{Pert.}$ in \cref{fig:triangle} (see \cref{sec:app_full_analysis}). In light of this, our framework could be explored with state-of-the-art CMB likelihoods where this anomaly is not present~\cite{Tristram:2023haj, Rosenberg:2022sdy, Naredo-Tuero:2024sgf}. The opposite effects of $\sum m_\nu^\mathrm{Backg.}$ and $\sum m_\nu^\mathrm{Pert.}$ on CMB lensing, together with the scale-dependence of perturbations effects, could also be leveraged to separate both effects in future high-precision determinations of CMB lensing~\cite{SPT-3G:2014dbx, SPT-3G:2024atg, LiteBIRD:2022cnt, LiteBIRD:2023iiy, SimonsObservatory:2018koc}. Moreover, some cosmological tensions are correlated with neutrino masses, motivating new studies that address them in the context of our separation of neutrino-mass effects. This may shed light on the physics that can alleviate these tensions. Beyond pure cosmology studies, our results provide a benchmark to build models that evade the cosmological neutrino-mass bound. In short, the main way to relax the cosmological bound is by modifying the expansion history of the Universe, with the ``kinematic'' properties of the model being less important. 

While we wait for a positive neutrino-mass signal from terrestrial experiments, cosmological measurements lead current limits. As they improve, we find ourselves in an era where cosmological limits are approaching values disfavored by oscillation experiments. If this tension grows, the solution may rely on a non-standard neutrino sector. If, in turn, a neutrino-mass signal is found, scrutinizing the robustness of this determination will be mandatory. In both cases, understanding the involved physical effects and degeneracies is key for solid progress in cosmology and particle physics. Future observations will guide the next steps for the physics of neutrinos, the first particle whose mass may be first measured outside laboratories.

\begin{acknowledgments}
This work has been supported by the Spanish MCIN/AEI/10.13039/501100011033 grants PID2020-113644GB-I00, PID2023-148162NB-C22 (RH and OM), PID2023-151418NB-I00 (RH), PID2021-123703NB-C21, PID2022-136510NB-C33 (IE), and PID2022-126224NB-C21 (TB and JS); and by the European Union’s Horizon 2020 research and innovation program under the Marie Skłodowska-Curie grants HORIZON-MSCA-2021-SE-01/101086085-ASYMMETRY and H2020-MSCA-ITN-2019/860881-HIDDeN. RH is supported by the Spanish grant FPU19/\allowbreak03348 of MU, and TB by the Spanish grant PRE2020-091896. The authors also acknowledge support from the Generalitat Valenciana grants PROMETEO/2019/083 and CIPROM/2022/69 as well as from COST Action CA21136 CosmoVerse (RH and OM). IE acknowledges support from the Basque Government (IT1628-22). The authors acknowledge the Galileo Galilei Institute (GGI) for Theoretical Physics (RH and OM), the Center for Cosmology and AstroParticle Physics (CCAPP) at the Ohio State University (RH), the Fermi National Accelerator Laboratory FERMILAB (TB, RH, OM and JS), the University of the Basque Country UPV/EHU (RH), the Institut de Ciències del Cosmos of the Universitat de Barcelona (RH), and Institut de Física Corpuscular (IE and TB) for their hospitality during the completion of this work. OM acknowledges the financial support from the MCIU with funding from the European Union NextGenerationEU (PRTR-C17.I01) and Generalitat Valenciana (ASFAE/2022/020). TB and JS acknowledge support from the ``Unit of Excellence Maria de Maeztu 2020-2023'' award to the ICC-UB CEX2019-000918-M. 
\end{acknowledgments}

\bibliographystyle{JHEP.bst}
\bibliography{biblio}

\clearpage
\onecolumngrid
\appendix
\renewcommand\thefigure{\thesection\arabic{figure}}

\setcounter{figure}{0}

\section{Evolution of neutrino perturbations}
\label{sec:nu_evo}

In this Appendix, we detail how we compute the effective neutrino sound-speed $c_\mathrm{eff}^2$ and the neutrino anisotropic stress $\sigma$, as functions of the perturbations neutrino mass $\sum m_\nu^\mathrm{Pert.}$. We then substitute their values in \cref{eq:delta_nu,eq:theta_nu}, that we solve numerically within the public code \texttt{CLASS} to compute the impact of $\sum m_\nu^\mathrm{Pert.}$ on CMB observables.

Technically, $c_\mathrm{eff}^2$ and $\sigma$ depend on $\sum m_\nu^\mathrm{Pert.}$ as well as on the gravitational potentials $\phi$ and $\psi$ (in this Appendix, as in the entire manuscript, we work in the Newtonian gauge). Physically, this captures the backreaction of gravity onto neutrinos, in the same way that the background equation of state $w_\nu$ depends both on the neutrino mass and on the scale factor $a$ (see \cref{eq:wnu}). To compute this dependence for massive, non-interacting neutrinos; we start from the collisionless Boltzmann equation for the perturbed neutrino distribution $\Psi$~\cite{Ma:1995ey},
\begin{equation}
  \frac{\partial \Psi}{\partial x} + i \mu \frac{q}{\epsilon} \Psi = \frac{\mathrm{d}\ln f_0}{\mathrm{d}\ln q} \left(i \mu \frac{\epsilon}{q} \psi - \frac{\mathrm{d}\phi}{\mathrm{d}x}\right) \, .
  \label{eq:Boltzmann_x}
\end{equation}
Here, $x \equiv k\eta$ is the product of Fourier wavenumber and conformal time, $q \equiv ap$ is the comoving 3-momentum with direction $\hat{n}$, $\mu \equiv \vec{k}\cdot\hat{n}$, $\epsilon \equiv \sqrt{q^2+m^2a^2}$ is the comoving energy, and $f_0$ is given by \cref{eq:f0} in the main text. Following Refs.~\cite{Shoji:2010hm,Nascimento:2023psl}, this equation can be implicitly solved in terms of $\phi$ and $\psi$. After expanding in Legendre polynomials and integrating by parts,
\begin{equation}\label{eq:implicit-Psi}
    \Psi_\ell(x) = \Psi(0) j_\ell\left( y(0, x)  \right)  + \frac{\mathrm{d}\ln f_0}{\mathrm{d}\ln q}  \left\lbrace\phi(0) j_\ell\left(y(0, x)\right) - \phi(x)j_\ell(0)  - \int_0^x \mathrm{d}x' \left[\frac{\epsilon}{q}\psi(x') + \frac{q}{\epsilon}  \phi(x')\right] j_\ell'\left(y(x', x)\right) \right\rbrace \, ,
\end{equation}
where 
\begin{equation}
\Psi(k, q, \mu, x) \equiv \sum_{\ell=0}^\infty (-i)^\ell (2\ell+1)\Psi_\ell(k, q, x) P_\ell(\mu) \, ,
\end{equation}
with $P_\ell$ the Legendre polynomials; $j_\ell(x)$ the spherical Bessel functions; and 
\begin{equation}
    y(x_1, x_2) \equiv \int_{x_1}^{x_2}\frac{q}{\epsilon(x')} \, \mathrm{d}x' \, ,
\end{equation}
the distance traveled by neutrinos between times $x_1/k$ and $x_2/k$, divided by the mode size. The super-horizon adiabatic initial condition for $\Psi$, assuming radiation domination and ultrarelativistic neutrinos, is~\cite{Ma:1995ey}
\begin{equation}
    \Psi(x=0) =  \frac{1}{2}\psi(\eta=0)\,  \frac{\mathrm{d}\ln f_0}{\mathrm{d}\ln q} \, .
\end{equation}

These equations allow to explicitly compute $\Psi(k, q, \mu, \eta)$ as a function of the gravitational potentials and the neutrino dispersion relation $\epsilon(q)$. The latter depends on the total perturbations mass $\sum m_\nu^\mathrm{Pert.}$, since $\epsilon(q) = \sqrt{q^2+a^2(m_\nu^\mathrm{Pert.})^2}$ with $m_\nu^\mathrm{Pert.}$ each individual perturbations neutrino mass. Once $\Psi$ is known, $c_\mathrm{eff}^2$ and $\sigma$ can be computed from \cref{eq:deltatower,eq:thetatower,eq:deltaPtower,eq:sheartower,eq:ceff_nu} in the main text.

We note that $\Psi$ as obtained in this Appendix could in principle be used to compute $\delta$ and $\theta$ and directly source the Einstein equations. This, however, would be inconsistent with energy-momentum conservation, \cref{eq:delta_nu,eq:theta_nu}, that enforces that the evolution of $\delta$ and $\theta$ depends on the background equation of state.

\clearpage \newpage

\section{Consistency with the background and gauge-invariance of the effective sound speed}
\label{sec:bkg-consistency}

As discussed in \cref{sec:pert} in the main text, the fact that adiabatic perturbations behave as background as $k \rightarrow 0$ may enforce a consistency relation between $c_s^2$, $\sigma$ and $w_\nu$. On top of that, an indeterminacy is introduced by $c_s^2$ being gauge-dependent. In this Appendix, we show that writing the perturbations equations \cref{eq:delta_nu,eq:theta_nu} in terms of the effective sound-speed $c_\mathrm{eff}^2$ overcomes both issues.

\subsection{Gauge invariance}
\label{sec:gauge}

Under a gauge transformation
\begin{align}
    \eta & \rightarrow \eta + \alpha(\vec{x}, \eta) \, ,\\
    \vec{x} & \rightarrow \vec{x} + \vec{\nabla} \beta(\vec{x}, \eta) + \vec{\epsilon}\,(\vec{x}, \eta) \, ,
\end{align}
matter perturbations transform as~\cite{Ma:1995ey}
\begin{align}
    \delta & \to \delta - \alpha \rho'/\rho \, , \label{eq:gauge_delta}\\
    \theta & \to \theta - \alpha k^2 \, , \label{eq:gauge_theta}\\
    \delta P & \to \delta P - \alpha P' \, , \label{eq:gauge_deltaP}\\
    \sigma & \to \sigma \, . \label{eq:gauge_sigma}
\end{align}

As a first consequence, $\sigma$ is gauge-invariant. Straightforward substitution of the gauge transformation in \cref{eq:ceff_nu} shows, using \cref{eq:ad}, that $c_\mathrm{eff}^2$ is also gauge-invariant.

\subsection{Consistency with background} \label{sec:app_bkg}

For adiabatic perturbations, super-horizon long-wavelength perturbations must behave as background at a slightly different time~\cite{Ma:1995ey,Lesgourgues:2013sjj,Dodelson:2020abc}. That is, 
\begin{align}
  \lim_{k\to 0}\,({P} + \delta P)(\eta) & ={P}(\eta + \delta \eta) = {P}(\eta) + P'(\eta) \delta \eta \, ,\\
  \lim_{k\to 0}\,(\rho + \delta\rho)(\eta) & ={\rho}(\eta + \delta \eta) = {\rho}(\eta) + \rho'(\eta) \delta \eta \, .
\end{align}
This immediately implies that, for our perturbations formalism to be consistent with the background,
\begin{equation}
  \lim_{k\to 0}\, c_s^2 \equiv \lim_{k\to 0}\,\frac{\delta P_\nu}{\delta \rho_\nu} = \frac{P'_\nu}{\rho'_\nu} = w_\nu - \frac{w'_\nu}{3\frac{a'}{a}(1+w_\nu)}\equiv c_{\mathrm{ad}}^2 \, 
\end{equation}
must be fulfilled. It is straightforward to check that, if this condition holds, enforcing the background equation \cref{eq:back_ev} at a conformal time $\eta + \delta \eta$ leads to \cref{eq:delta_nu} (assuming $\theta \to 0$, which follows from \cref{eq:thetatower} as $k\to 0$).

To write this condition as a condition on $c_\mathrm{eff}$, we invert~\cref{eq:ceff_nu}
\begin{equation}
  c_s^2 = c_\mathrm{eff}^2 + (c_\mathrm{eff}^2 - c_\mathrm{ad}^2) \frac{3 \frac{a'}{a}(1+w_\nu) \theta}{k^2 \delta} \, ,
\end{equation}
It follows from this equation that, if
\begin{equation} \label{eq:consistency_ad}
    \lim_{k\to 0} \frac{3 \frac{a'}{a}(1+w_\nu) \theta}{k^2 \delta} = -1
\end{equation}
\emph{consistency with background is satisfied for all values of $c_\mathrm{eff}^2$ and $w_\nu$}. One can check (see \cref{eq:gauge_delta,eq:gauge_deltaP,eq:gauge_theta,eq:gauge_sigma}) that this condition is gauge-invariant.

We now turn to explicitly checking that \cref{eq:consistency_ad} is indeed satisfied for adiabatic perturbations. In the conformal Newtonian gauge, super-horizon adiabatic initial conditions in the radiation-domination regime for ultrarelativistic neutrinos correspond to $\delta = -2\psi$, $\theta = k^2\eta\psi/2$, $w_\nu = 1/3$, $a'/a = \eta^{-1}$~\cite{Ma:1995ey}; which satisfy \cref{eq:consistency_ad}. \Cref{eq:consistency_ad} is also maintained by super-horizon evolution: using \cref{eq:delta_nu,eq:theta_nu} with $k\to 0$ (where $\theta \to 0$ and $\sigma\to 0$, see \cref{eq:thetatower,eq:sheartower}), as well as the Friedmann and perturbed Einstein equations~\cite{Ma:1995ey}, we obtain
\begin{equation}
  \frac{\mathrm{d}}{\mathrm{d}\eta} \left[\frac{3 \frac{a'}{a}  (1+w_\nu) \theta}{k^2} + \delta \right] = \frac{a'}{a}  \left[-\frac{3}{2}  \left(1 + \frac{{P}_\mathrm{tot}}{{\rho}_\mathrm{tot}}\right)+ 3w_\nu\right]\left[\frac{3\frac{a'}{a} (1+w_\nu)\theta}{k^2} + \delta \right] \, , \\
\end{equation}
with $P_\mathrm{tot}$ and $\rho_\mathrm{tot}$ the total pressure and energy density of the Universe. Since the right-hand side vanishes for the initial conditions, super-horizon evolution keeps the identity true at all times, even if $\frac{a'}{a}$  or $w_\nu$ change with time.

Finally, our definition of $\sigma$ is also consistent with the background. Adiabatic perturbations behave as background at super-horizon scales, which implies a diagonal stress-energy tensor, i.e., $\sigma\to 0$ as $k\to 0$. \Cref{eq:implicit-Psi,eq:sheartower} trivially fulfill this, because, $j_2(0)=0$. That is, the anisotropic stress is non-zero only when the distance travelled by neutrinos is of the order of the mode size. Since $\sigma$ is a gauge-invariant quantity, this is true in all gauges.

\clearpage \newpage

\section{Detailed physical effects on the CMB anisotropies} \label{sec:app_CMB}

In this Appendix, we provide more details and explicit checks of the different effects of $\sum m_\nu^\mathrm{Backg.}$ and $\sum m_\nu^\mathrm{Pert.}$ on the CMB anisotropies; with an emphasis on the underlying physical processes.

Unless stated otherwise, all the CMB power spectra shown throughout this manuscript are obtained by fixing the standard $\Lambda$CDM parameters to their best-fit values of the Planck 2018 $\mathrm{TT,TE,EE+lowE+lensing}$ analysis~\cite{Planck:2018vyg}. The values of these parameters are $100\theta_s = 1.04172,\,\omega_b = 0.02237,\,\omega_{\mathrm{cdm}} = 0.1200,\,\log(10^{10}A_s) = 3.044,\,n_s = 0.9649,\, \tau_{\mathrm{reio}} = 0.0544$, and $\sum m_\nu = 0.06~\mathrm{eV}$.

\subsection{Modified Sachs-Wolfe separation}

The amplitude of the $\ell$-th multipole of CMB temperature anisotropies, without including CMB lensing, is given by
\begin{equation}
  C_\ell = \frac{1}{2\pi^2} \int \frac{\mathrm{d}k}{k} \mathcal{P}_\mathcal{R}(k) [\Theta_{\gamma \ell}(\eta_0, k)]^2 \, ,
\end{equation}
with $\mathcal{P}_\mathcal{R}(k)$ the primordial curvature spectrum---in the standard case $\mathcal{P}_\mathcal{R} = A_s (k/k_0)^{n_s-1}$~\cite{Lesgourgues:2013sjj}---, and $\Theta_{\gamma \ell}(\eta_0, k)$ the present photon temperature transfer function, normalized so that the curvature perturbation at initial times $\mathcal{R} = 1$. 

To gain physical intuition, it is customary to write $\Theta_{\gamma \ell}$ in terms of a line-of-sight integral~\cite{Zaldarriaga:1995gi}
\begin{equation}\label{eq:los}
    \Theta_{\gamma \ell}\left(\eta_0, k\right)=\int_{\eta_{\text {in }}}^{\eta_0} \left\{g(\eta)\left(\Theta_{\gamma 0}+\psi\right)+\left[g(\eta) k^{-2} \theta_{\mathrm{B}}\right]^{\prime}+e^{-\tau}\left(\phi^{\prime} +\psi^{\prime}\right)\right\} j_\ell\left[k\left(\eta_0-\eta\right)\right]\mathrm{d} \eta\, ,
\end{equation}
with $\eta_\mathrm{in}$ the initial conformal time, $\theta_B$ the baryon velocity divergence,
\begin{equation}
  \tau(\eta) \equiv \int_{\eta}^{\eta_0}  a(\eta') n_e(\eta') \sigma_\mathrm{T}\,\mathrm{d}\eta'
\end{equation}
the photon optical depth, which changes from 0 well before recombination to 1 after recombination; and $g(\eta) \equiv [e^{-\tau(\eta)}]'$ the visibility function, which is sharply peaked around recombination~\cite{Lesgourgues:2013sjj}. The Bessel function $j_\ell$ sets $\ell \sim k (\eta_0-\eta)$.

\Cref{eq:los} provides a simple physical interpretation of the observed CMB anisotropies. The first term can be understood as the temperature anisotropy at recombination, redshifted by the local gravitational potential (Sachs-Wolfe effect, SW); the second term as a Doppler shift due to baryon velocities at recombination; and the third term as the accumulated gravitational redshift as CMB photons travels from recombination to us (Integrated Sachs-Wolfe effect, ISW). The ISW effect can be further understood as caused by the depth of gravitational potential wells changing while photons are inside them, due to the combined action of structure growth and the expansion of the Universe.

This separation, however, is not particularly appropriate to study the impact of neutrinos. Both $(\Theta_{\gamma 0} + \psi)$ and $(\phi' + \psi')$ depend on the neutrino anisotropic stress $\sigma$ even at super-horizon scales~\cite{Ma:1995ey}. Thus, the SW contribution to modes that are much larger than the horizon would depend on $\sigma$; and, as $\sigma$ changes with time, there would be an ISW effect even for modes that are well outside the horizon. This would contradict the expected super-horizon behavior of adiabatic perturbations: as $k \to 0$, these perturbations behave as background and their evolution cannot depend on perturbation-related quantities such as $\sigma$.

Indeed, one can numerically check that the $\ell \rightarrow 0$ limit of \cref{eq:los} is independent of the anisotropic stress around recombination, even if the SW and ISW terms separately are not. Technically, the SW term can be integrated by parts, leading to terms proportional to $e^{-\tau}$ that may as well be interpreted as an ISW contribution. Physically, as argued by Bardeen in his seminal paper on gauge-invariant cosmological perturbations~\cite{Bardeen:1980kt}, there is an ambiguity of what one means by a temperature or metric perturbation at scales comparable to or bigger than the horizon. 

To facilitate physical understanding, below we redefine the separation between SW (understood as the temperature fluctuation at recombination, redshifted by the local gravitational potential) and ISW (understood as the accumulated gravitational redshift from recombination until today) contributions, so that they both depend only on background quantities in the $k \to 0$ limit.

To such purpose, we split the last term in \cref{eq:los} as
\begin{equation}\label{eq:phipsi-sep}
    \phi + \psi = \frac{6(1+w_\mathrm{tot})}{5+3w_\mathrm{tot}} \left[ \phi + \frac{2}{3(1+w_\mathrm{tot})} \left( \psi + \frac{a}{a'}\phi'\right)\right]
    - \frac{1+3w_\mathrm{tot}}{5+3w_\mathrm{tot}} \left( \phi - \psi + \frac{2}{3}\frac{a}{a'}\phi'\frac{6}{1+3w_\mathrm{tot}}\right) \, ,
\end{equation}
with $w_\mathrm{tot} \equiv P_\mathrm{tot}/\rho_\mathrm{tot}$ the total equation of state of the Universe. We identify the comoving curvature perturbation
\begin{equation}
  \mathcal{R} \equiv \phi + \frac{a'}{a}\frac{\theta_\mathrm{tot}}{k^2} = \phi + \frac{2}{3(1+w_\mathrm{tot})}\left(\psi + \frac{a}{a'}\phi'\right) \, ,
\end{equation} 
with $\theta_\mathrm{tot}$ the total velocity divergence of all the matter content in the Universe. Here, we have used the Einstein equations to express $\theta_\mathrm{tot}$ in terms of $\phi$ and $\psi$. Thus, we can rewrite \cref{eq:phipsi-sep} as
\begin{equation}\label{eq:phipsi-sep-2}
    \phi + \psi = \frac{6(1+w_\mathrm{tot})}{5+3w_\mathrm{tot}} \mathcal{R} - \frac{1+3w_\mathrm{tot}}{5+3w_\mathrm{tot}} \left( \phi - \psi + \frac{2}{3}\frac{a}{a'}\phi'\frac{6}{1+3w_\mathrm{tot}}\right)  \, .
\end{equation}
It immediately follows from the Einstein equations that $\mathcal{R}$ is conserved on super-horizon scales.

Substituting \cref{eq:phipsi-sep-2} in the third term in \cref{eq:los}, and integrating by parts the second term in \cref{eq:phipsi-sep-2}, we obtain
\begin{equation}
    \Theta_{\gamma \ell}\left(\eta_0, k\right)=\int_{\eta_{\text {in }}}^{\eta_0} \left\{g(\eta) \Theta_{\gamma \ell}^\mathrm{SW}+\left[g(\eta) k^{-2} \theta_{\mathrm{B}}\right]^{\prime}j_\ell\left[k\left(\eta_0-\eta\right)\right]+e^{-\tau}\Theta_{\gamma \ell}^\mathrm{ISW}\right\} \mathrm{d} \eta\, ,
\end{equation}
with
\begin{equation}\label{eq:our-sw}
\begin{split}
\Theta_{\gamma \ell}^\mathrm{SW}  &\equiv 
                        \left[\Theta_{\gamma 0} + \psi +\frac{1+3w_\mathrm{tot}}{5+3w_\mathrm{tot}}\left(\phi - \psi + \frac{2}{3}\frac{a}{a'}\phi' \frac{6}{1+3w_\mathrm{tot}}\right)
                         \right]j_\ell\left[k\left(\eta_0-\eta\right)\right] \, = \\ 
                         &=
                        \left[\frac{1+3w_{\mathrm{tot}}}{5+3w_{\mathrm{tot}}}\mathcal{R}+ 
                               \left(\Theta_{\gamma 0}+
                                     \frac{a'}{a}\frac{\theta_{\mathrm{tot}}}{k^2}\right)
                         \right]j_\ell\left[k\left(\eta_0-\eta\right)\right] \, ,
\end{split}
\end{equation}
where we have used the Einstein equations to go from the first to the second line; and
\begin{equation}\label{eq:our-isw}
\Theta_{\gamma\ell}^\mathrm{ISW} \equiv 
                        \left(\frac{6(1+w_{\mathrm{tot}})}{5+3w_{\mathrm{tot}}}\mathcal{R}\right)' j_\ell[k(\eta_0-\eta)] 
                       - \frac{1+3 w_\mathrm{tot}}{5+3w_\mathrm{tot}}\left(\phi-\psi + \frac{4}{1+3w_{\mathrm{tot}}}\frac{a}{a'}\phi'\right)
                        k\, j_\ell'[k(\eta_0-\eta)]\, .
\end{equation}

The first term in \cref{eq:our-sw} is, on super-horizon scales, explicitly $\sigma$-independent (we normalize the transfer function to $\mathcal{R}=1$ at initial times). Since on such scales $\mathcal{R}$ is conserved, it only depends on the background quantity $w_\mathrm{tot}$---the amplitude of super-horizon perturbations does depend on the background equation of state~\cite{Kodama:1984ziu, Mukhanov_2005}---, and it provides the correct $\ell \to 0$ SW plateau of the CMB power spectrum~\cite{Sachs:1967er, Dodelson:2020abc} (see \cref{eq:SW_plateau} below). The second term in \cref{eq:our-sw} is zero for super-horizon scales and adiabatic perturbations (see \cref{sec:app_bkg}). It corresponds to one of the gauge-invariant density perturbations defined by Bardeen~\cite{Bardeen:1980kt}, and it represents the photon temperature perturbations in a gauge comoving with matter.

The first term in \cref{eq:our-isw} is also, on super-horizon scales, explicitly $\sigma$-independent. It only introduces an ISW effect if $w_\mathrm{tot}$ changes, which would affect the super-horizon gravitational potentials. The second term in \cref{eq:our-isw} is explicitly suppressed as $k \to 0$.

\begin{figure}
    \centering
    \includegraphics[width=0.495\linewidth]{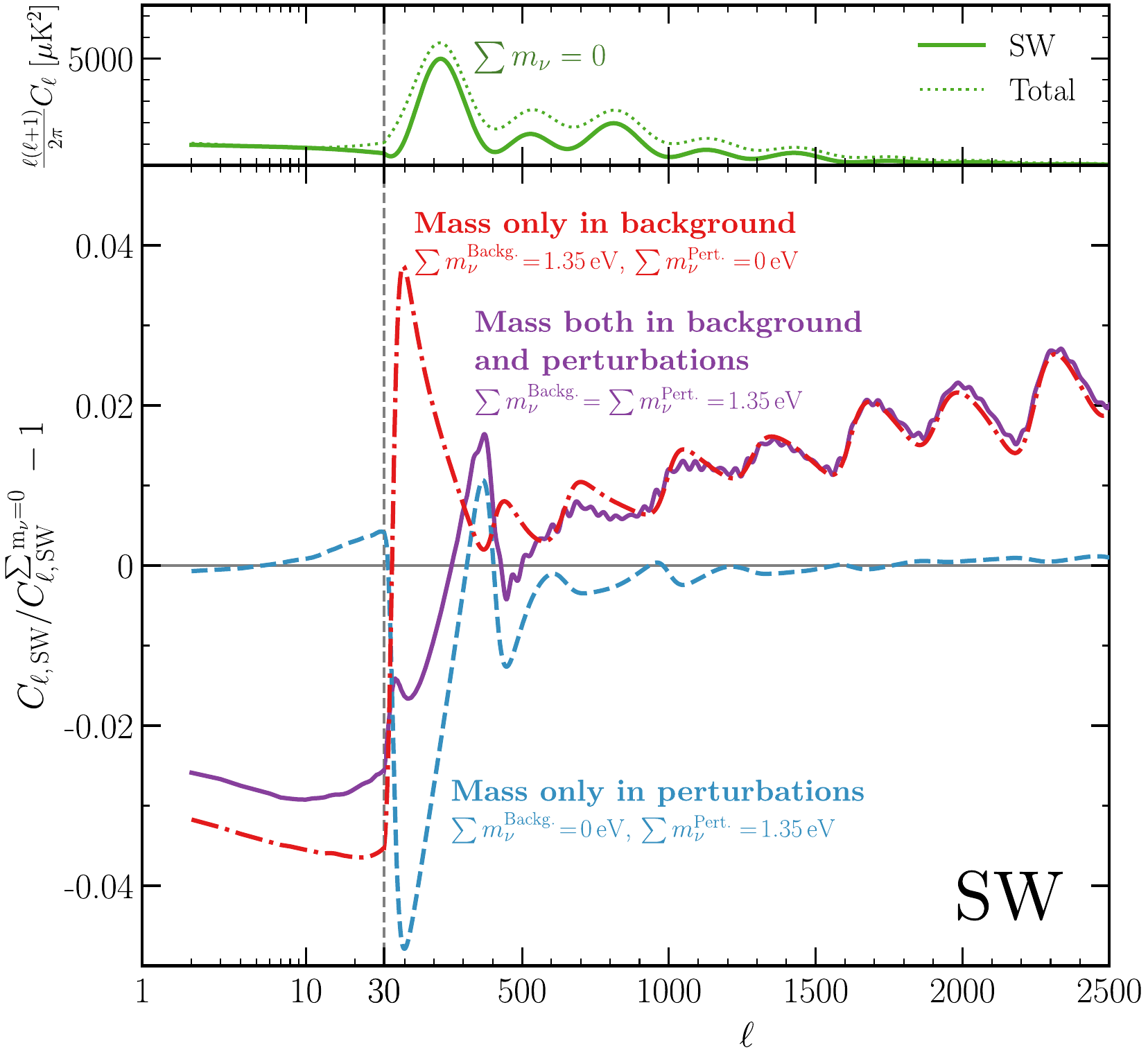}
    \includegraphics[width=0.495\linewidth]{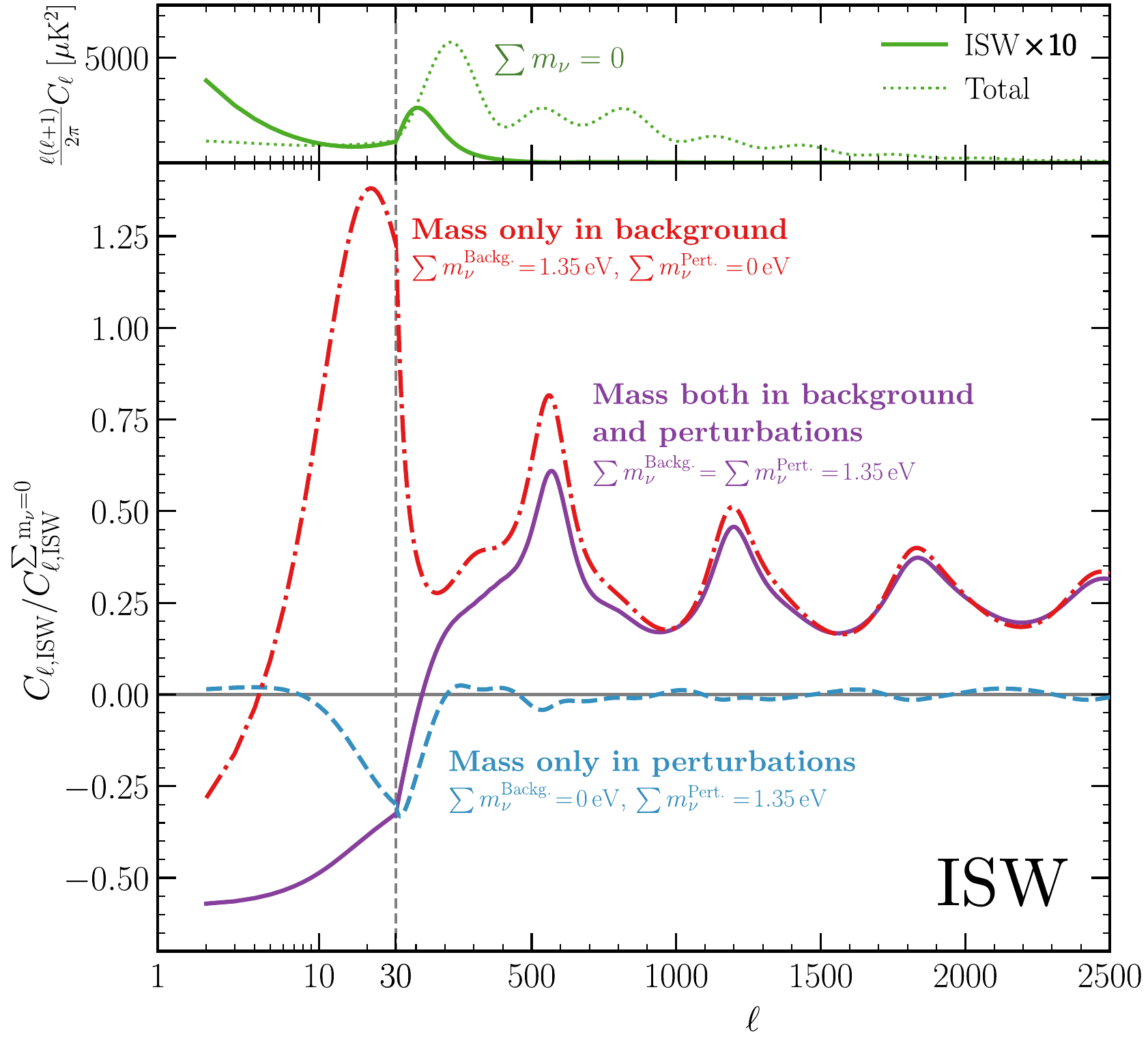}
    \caption{Background- and perturbations-induced impact of $\sum m_\nu$ on the Sachs-Wolfe (SW, left) and Integrated Sachs-Wolfe (ISW, right) effects as defined in~\cref{eq:our-sw,eq:our-isw}, without including CMB lensing. The SW contribution can be understood as due to the temperature fluctuations at recombination, redshifted by the local gravitational potential. In turn, the ISW contribution can be understood as the accumulated gravitational redshift from recombination until today. Top plots show the contribution to the total CMB anisotropy power spectrum, and bottom plots the relative difference with respect to massless neutrinos. \textit{The SW contribution, as defined in this work, only depends on background quantities as $k \to 0$.}}
    \label{fig:SW-ISW}
\end{figure}

\Cref{fig:SW-ISW} shows our separation among SW and ISW contributions to the CMB temperature anisotropies, for different values of $\sum m_\nu^\mathrm{Backg.}$ and $\sum m_\nu^\mathrm{Pert.}$. The blue line has the same background as $\sum m_\nu = 0$, and the SW contribution is consequently suppressed as $\ell \rightarrow 0$. The purple and red lines also share the same background evolution, but they have sub-percent differences as $\ell \rightarrow 0$. This is reasonable, as these multipoles correspond to scales where $(k_{\ell=2}\eta_\mathrm{rec})^2\sim 0.004$, so they are not completely out of the horizon around recombination.

\subsection{Physical effects on the SW contribution}\label{sec:app_cmb_sw}

The SW effect accounts for temperature anisotropies at recombination, redshifted by the local gravitational potential. We now break down the different physical effects of $\sum m_\nu^\mathrm{Backg.}$ and $\sum m_\nu^\mathrm{Pert.}$ on the SW contribution in~\cref{fig:SW-ISW}. 

\subsubsection{Low multipoles}

Low multipoles ($\ell \lesssim 10$) correspond to modes that are super-horizon at recombination. Their evolution only depends on background quantities, and it can be analytically estimated. Setting $k\ll \frac{a'}{a}$ in \cref{eq:our-sw}, the $\ell \to 0$ plateau is given by~\cite{Dodelson:2020abc}
\begin{equation}
    \lim_{\ell \to 0}\ell(\ell+1)C_\ell^{\mathrm{SW}} \approx 8\left[\frac{1+3w_\mathrm{tot}(\eta_{\mathrm{rec}})}{5+3w_\mathrm{tot}(\eta_{\mathrm{rec}})}\right]^2 A_s\, , \label{eq:SW_plateau}
\end{equation}
which coincides with the standard result $\ell(\ell+1)C_\ell^{\mathrm{SW}}\approx 8A_s/25$ for matter domination, $w_{\mathrm{tot}}(\eta_{\mathrm{rec}}) = 0$~\cite{Sachs:1967er,Dodelson:2020abc}. Increasing $\sum m_\nu^\mathrm{Backg.}$ reduces the neutrino contribution to $w_\mathrm{tot}$, reducing the $\ell \to 0$ limit of the SW contribution. There is also a sub-percent effect of $\sum m_\nu^\mathrm{Pert.}$ that, as explained above, is due to the corresponding modes not being completely out of the horizon at recombination.

\subsubsection{Intermediate multipoles}

Intermediate multipoles ($10 \lesssim \ell \lesssim 500$) correspond to modes that are comparable to the horizon at recombination. Their evolution is largely influenced by the gravitational potentials $\phi$ and $\psi$. It can be approximated as~\cite{Hu:1995kot}
\begin{equation} \label{eq:theta_gamma_oscillator}
    m_{\mathrm{eff}}\Theta_{\gamma 0}'' + k^2 \frac{\Theta_{\gamma 0}}{3}\simeq - m_{\mathrm{eff}}\left(k^2\frac{\psi}{3}-\phi''\right)\, ,
\end{equation}
where $m_{\mathrm{eff}} \equiv 1 + \frac{\rho_b + P_b}{\rho_\gamma + P_\gamma}\simeq 1+ 3\rho_b/(4\rho_\gamma)$; with $\rho_b$, $P_b$, $\rho_\gamma$, and $P_\gamma$ the baryon energy density, baryon pressure, photon energy density, and photon pressure, respectively. \Cref{eq:theta_gamma_oscillator} is the equation of a forced harmonic oscillator, with the gravitational potentials playing the role of an external force.

\begin{figure}
    \centering
    \includegraphics[width=0.95\linewidth,clip,trim={0 15pt 0 0}]{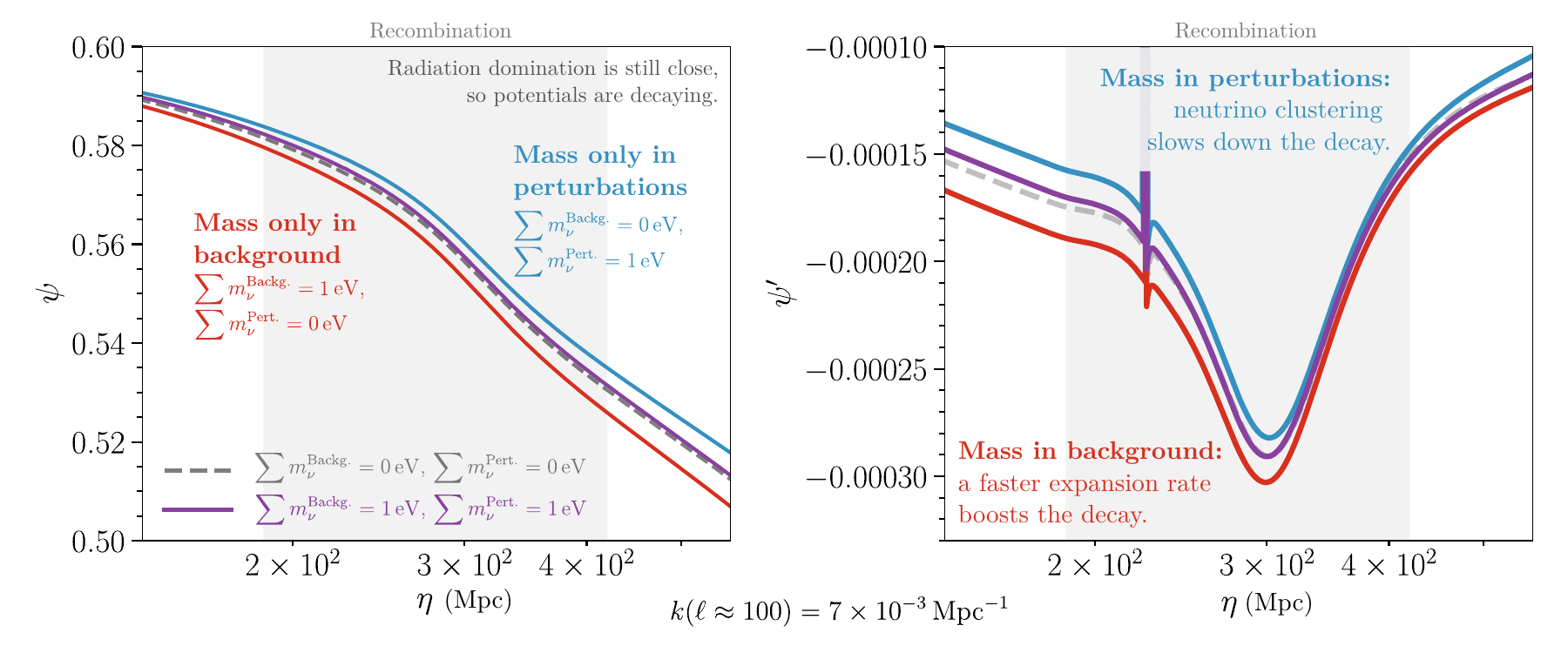}
    \caption{Gravitational potential $\psi$ (left) and its derivative with respect to conformal time (right) for a mode entering the horizon ($k\eta_{\mathrm{rec}}\sim 1$) at the time of recombination. We normalize to an initial comoving curvature perturbation $\mathcal{R} = 1$. This mode corresponds to $\ell \sim 100$. \textit{Background and perturbations neutrino masses affect intermediate-$\ell$ CMB anisotropies by boosting and slowing down the decay of $\psi$, respectively.}}
    \label{fig:psi-decay}
\end{figure}

Decaying gravitational potentials enhance the amplitude of oscillations via \cref{eq:theta_gamma_oscillator}. Physically, large initial potentials force the fluid into a highly compressed state. If they then decay, photon pressure overcomes gravity and the photon-baryon fluid oscillates with a larger amplitude. On top of that, decaying gravitational potentials reduce the redshift experienced by CMB photons as they leave the last-scattering surface, further increasing the anisotropies. For a detailed explanation of these effects, we refer to the work of Ref.~\cite{Hu:1995kot}. Gravitational potentials decay in the radiation-dominated era~\cite{Dodelson:2020abc}, and since recombination happens soon after matter-radiation equality, they are still decaying when SW anisotropies get frozen (see \cref{fig:psi-decay}).

The effect of neutrino masses is then straightforward. A large $\sum m_\nu^\mathrm{Backg.}$ increases the expansion rate of the Universe, boosting the decay of gravitational potentials, as shown in \cref{fig:psi-decay}; and increasing the SW contribution to CMB anisotropies, as shown in \cref{fig:SW-ISW}. A large $\sum m_\nu^\mathrm{Pert.}$ enhances neutrino clustering above the free-streaming scale, slowing down the decay of gravitational potentials, as shown in \cref{fig:psi-decay}; and decreasing the SW contribution to CMB anisotropies, as shown in \cref{fig:SW-ISW}. These effects are opposite, and thus the net effect for standard massive neutrinos with $\sum m_\nu^\mathrm{Backg.} = \sum m_\nu^\mathrm{Pert.}$ is smaller, as shown in \cref{fig:psi-decay,fig:SW-ISW}.

\begin{figure}
    \centering
    \includegraphics[width=0.495\linewidth]{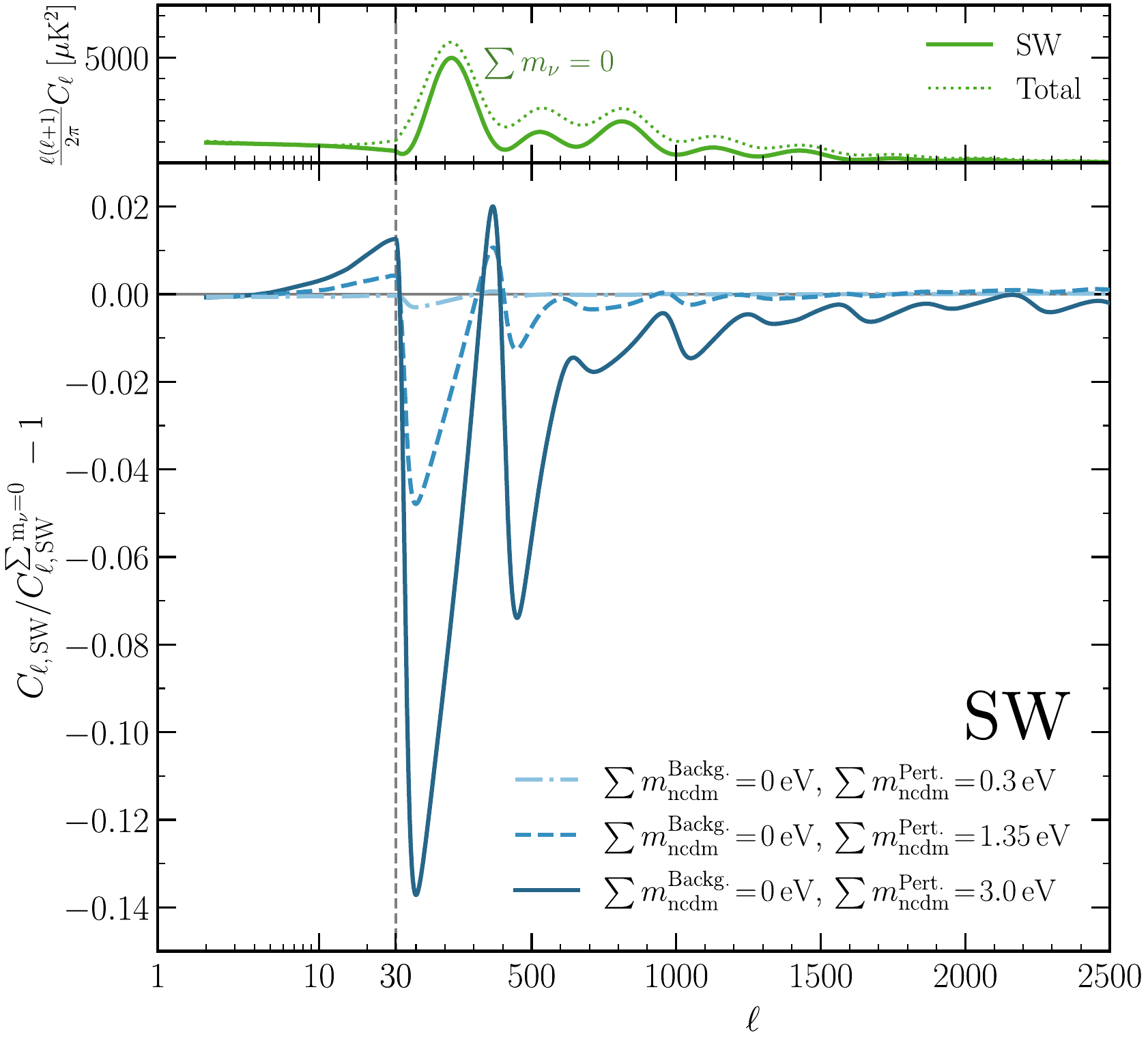}
    \caption{Scale-dependent impact of increasing $\sum m_\nu^{\mathrm{Pert.}}$ on the SW contribution to CMB anisotropies. Larger $\sum m_\nu^{\mathrm{Pert.}}$ slow down the decay of gravitational potentials through neutrino clustering (see~\cref{fig:psi-decay}), reducing the amplitude of acoustic oscillations at recombination. Since neutrino clustering is a scale-dependent effect controlled by the free-streaming length, $k_{\mathrm{FS}}$, \cref{eq:k_FS}, \emph{larger perturbations masses increase $k_{\mathrm{FS}}$ and propagate the impact to higher multipoles}.}
    \label{fig:SW-3eV}
\end{figure}

\Cref{fig:SW-3eV} shows the scale dependence of the clustering effect induced by $\sum m_\nu^\mathrm{Pert.}$. As described in \cref{sec:pert} in the main text, neutrino clustering is a scale-dependent ``kinematic'' effect that reflects that neutrinos do not move at the speed of light. The characteristic scale, $k_{\mathrm{FS}}$, is proportional to $\sum m_\nu^{\mathrm{Pert.}}$, see \cref{eq:k_FS}. As can be seen in \cref{fig:SW-3eV}, if $\sum m_\nu^{\mathrm{Pert.}}$ increases, the aforementioned depletion of CMB anisotropies affects higher multipoles.

\subsubsection{High multipoles}

High multipoles ($\ell \gtrsim 500$) correspond to modes which enter the horizon much before recombination. As discussed in \cref{sec:CMB} in the main text, these modes are affected by diffusion damping before recombination, an SW effect controlled by the damping scale $\theta_D$, \cref{eq:theta-damping}. Since $\theta_D$ only depends on background quantities, the high-$\ell$ contribution in \cref{fig:SW-ISW} depends only on $\sum m_\nu^{\mathrm{Backg.}}$ and not on $\sum m_\nu^{\mathrm{Pert.}}$. An increased $\sum m_\nu^{\mathrm{Backg.}}$ boosts the expansion rate of the Universe, reducing $\theta_D$ and enhancing the high-$\ell$ anisotropies in \cref{fig:SW-ISW}. We have checked that all the enhancement is due to a modified $\theta_D$, because it can be completely removed by artificially keeping $\theta_D$ fixed (technically, this is achieved by changing the primordial Helium fraction $Y_p$, which rescales $n_e$ in \cref{eq:theta-damping}~\cite{Lesgourgues:2013sjj}).

\subsection{Physical effects on the ISW contribution}

The ISW contribution accounts for the accumulated gravitational redshift of CMB photons on their way to Earth. Physically, when photons enter gravitational potential wells they get redshifted, and when they exit them they get blueshifted. If the depth of the gravitational potential wells is constant, both effects compensate each other. A nonzero ISW effect thus requires time-dependent gravitational potentials. This can only be the case if the Universe is \emph{not} matter-dominated~\cite{Dodelson:2020abc}.

The ISW effect is only relevant at multipoles $\ell \lesssim 200$, as shown in the top right panel of \cref{fig:SW-ISW}. At smaller scales, gravitational potentials experience a long period of radiation domination inside the horizon, and they are suppressed. It can be split into an early and a late contribution. 

The early contribution is due to the residual radiation left after recombination, and it affects modes that were inside the horizon at that time, $10 \lesssim \ell \lesssim 200$.  As discussed above (see \cref{fig:psi-decay}), background neutrino masses boost the decay of gravitational potentials and perturbations masses slow it down. Therefore, $\sum m_\nu^\mathrm{Backg.}$ and $\sum m_\nu^\mathrm{Pert.}$ produce an enhancement and a depletion of the ISW effect, respectively; as can be seen in \cref{fig:SW-ISW}.

The late contribution is due to the late-time epoch of accelerated expansion caused by the cosmological constant $\Lambda$, and it affects modes that entered the horizon recently, $\ell \lesssim 10$. $\Lambda$ makes the gravitational potentials decay, inducing a late ISW effect. As explained in \cref{sec:CMB} in the main text, the main neutrino-mass effect is indirect: the angular scale of CMB peaks $\theta_s$, \cref{eq:theta_s}, is very well measured, and $\sum m_\nu^\mathrm{Backg.}$ would modify it by changing the expansion rate of the Universe. To compensate for this effect, $H_0$ gets reduced, reducing $\Lambda$ and the late ISW effect as can be seen in \cref{fig:SW-ISW}. There is also a subleading reduction caused by $\sum m_\nu^\mathrm{Pert.}$, as neutrino clustering slows down the $\Lambda$-induced decay of potentials. This effect is small, see \cref{fig:cmb1,fig:SW-ISW}, and it is only present if $\sum m_\nu^{\mathrm{Backg.}} \neq 0$, because otherwise the neutrino energy density is too diluted at late times and neutrinos do not affect gravitational potentials.

\subsection{CMB lensing} \label{sec:app_lensing}

On their way to Earth, CMB photons get randomly deflected by the gravitational pull of the large-scale structure of the Universe~\cite{Dodelson:2020abc,Lewis:2006fu}. This weak gravitational lensing is a second-order non-linear effect, since it is a perturbative deflection of perturbative anisotropies, but with the precision of current data it is detectable. Lensing has two observable features on CMB anisotropies, that are more prominent at high multipoles. First, it smoothes out the power spectrum, since it mixes photons coming from different points in the last scattering surface. Second, it transfers power from large scales to small scales, due to its non-linear nature. This leads to increased anisotropies at high $\ell$.

Neutrino masses affect the evolution of gravitational potentials, as explained above and shown in~\cref{fig:psi-decay}. Consequently, they modify CMB lensing. On the one hand, background neutrino masses increase the expansion rate of the Universe and boost the decay of gravitational potentials. This reduces CMB lensing. On the other hand, perturbations masses make neutrinos cluster below the free-streaming scale and slow down the decay of potentials. This enhances CMB lensing in a scale-dependent way. 

\begin{figure}[b]
    \centering
    \includegraphics[width=0.495\linewidth]{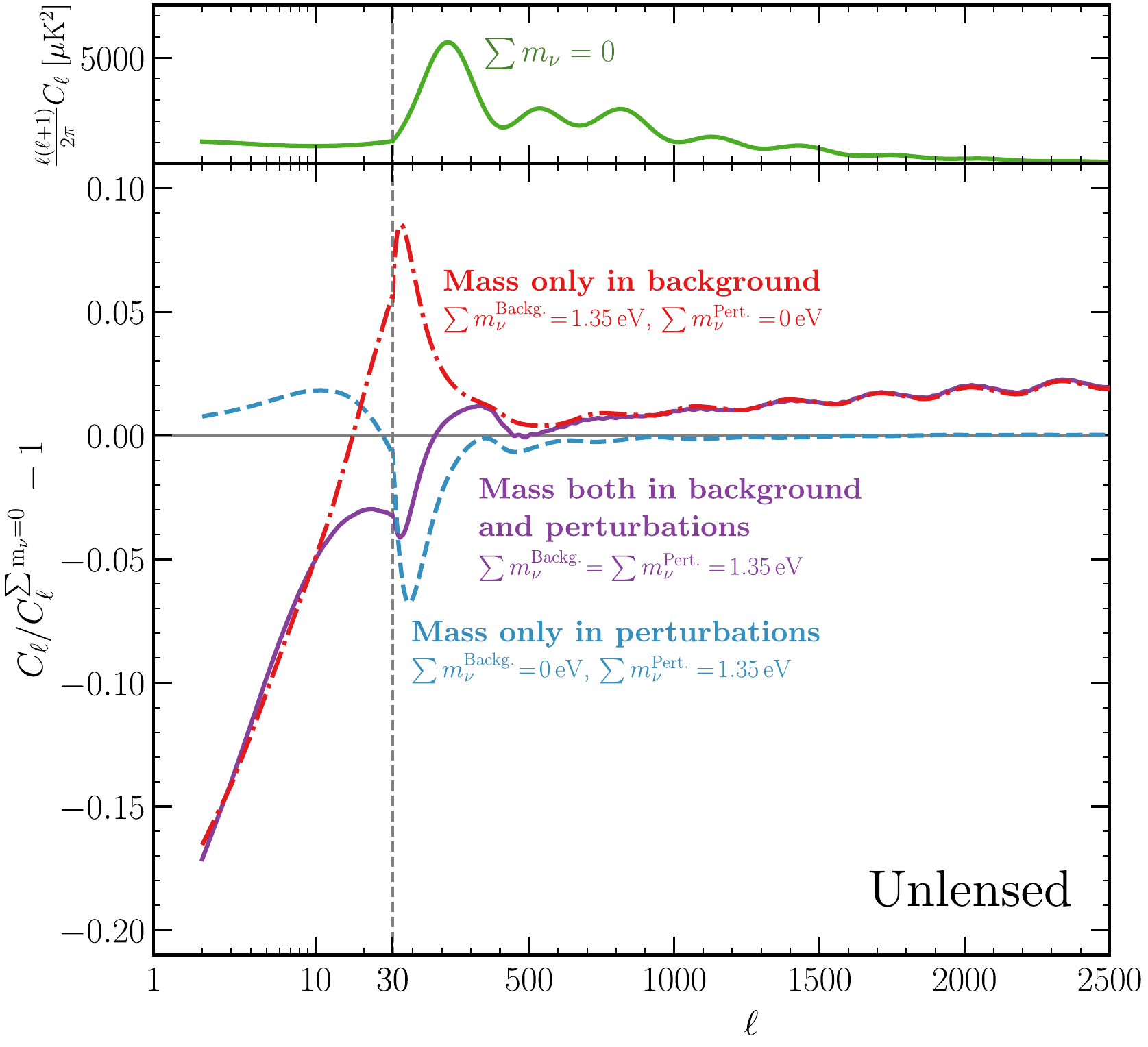}
    \includegraphics[width=0.495\linewidth]{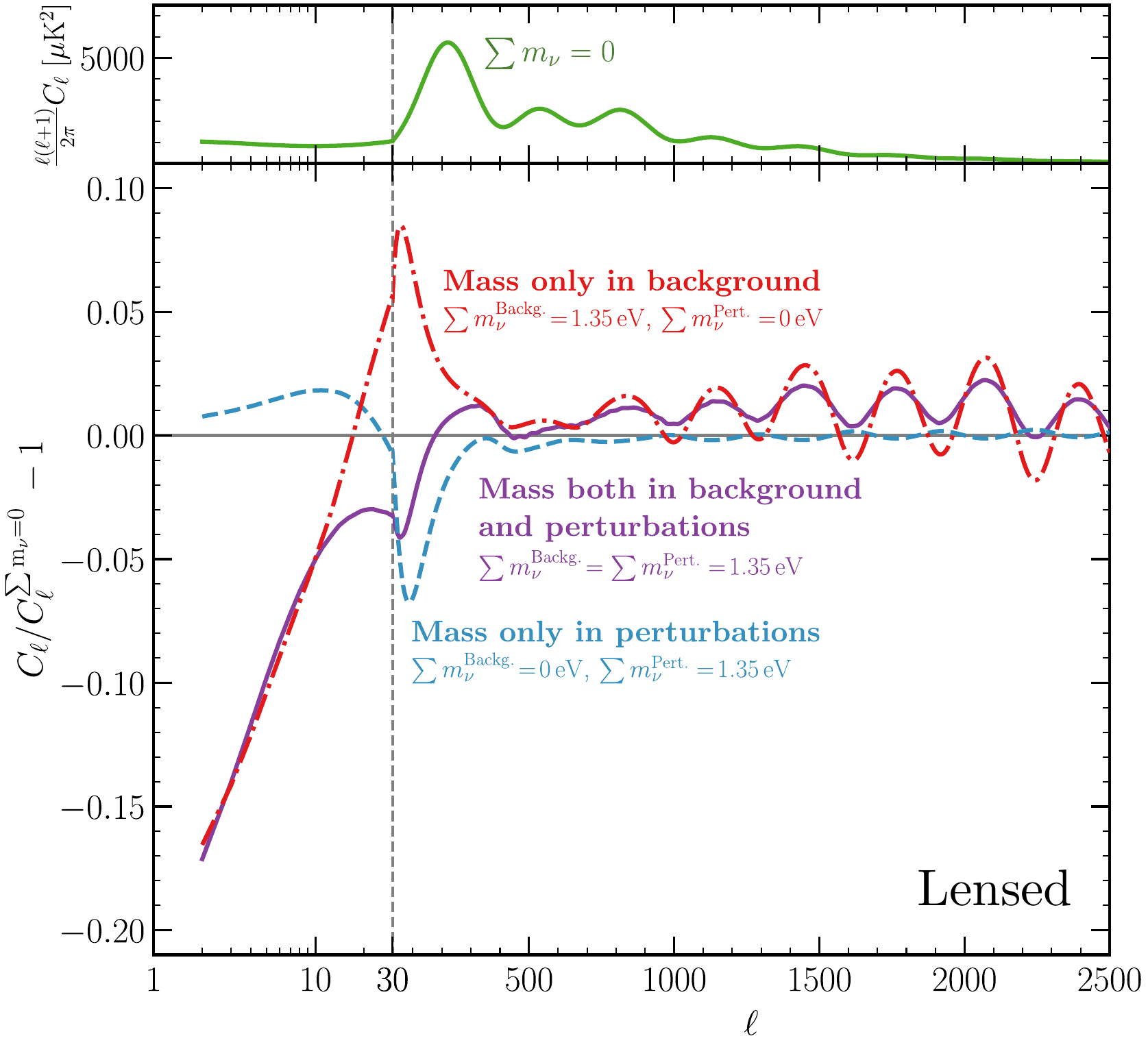}
    \caption{Impact of $\sum m_\nu^{\mathrm{Backg.}}$ and $\sum m_\nu^{\mathrm{Pert.}}$ on CMB anisotropies, with (left) and without (right) the effect of gravitational lensing. At high $\ell$, unlensed anisotropies only depend on $\sum m_\nu^{\mathrm{Backg.}}$ through diffusion damping. However, both $\sum m_\nu^{\mathrm{Backg.}}$ and $\sum m_\nu^{\mathrm{Pert.}}$ have an effect on the lensed anisotropies, and they do so in opposite directions. \textit{$\sum m_\nu^{\mathrm{Backg.}}$ reduces CMB lensing, while $\sum m_\nu^{\mathrm{Pert.}}$ enhances it.}}

    \label{fig:cmb-lensing-effects}
\end{figure}

\Cref{fig:cmb-lensing-effects} shows how $\sum m_\nu^{\mathrm{Backg.}}$ and $\sum m_\nu^{\mathrm{Pert.}}$ control these effects. The left panel shows CMB anisotropies without lensing, while the right panel includes lensing. The high-$\ell$ tail of the unlensed power spectrum only depends on $\sum m_\nu^{\mathrm{Backg.}}$ through diffusion damping, as explained in~\cref{sec:app_cmb_sw}. In turn, since $\sum m_\nu^{\mathrm{Backg.}}$ and $\sum m_\nu^{\mathrm{Pert.}}$ have opposite effects on lensing, they produce opposite features at lensed high-$\ell$. As $\sum m_\nu^{\mathrm{Backg.}}$ reduces lensing, CMB anisotropies are less smoothed, which is visible as wiggles in phase with the CMB peaks. Weaker lensing also implies less power transferred to high multipoles, which leads to a slight depletion at $\ell \gtrsim 2000$. As $\sum m_\nu^{\mathrm{Pert.}}$ increases lensing, its effect is opposite, reducing the amplitude of the wiggles and slightly enhancing anisotropies in the $\ell \gtrsim 2000$ region. Perturbations mass effects are only relevant if $\sum m_\nu^{\mathrm{Backg.}}\neq0$, because otherwise the neutrino energy density is too diluted at late times and neutrinos do not affect gravitational potentials.

\begin{figure}[b]
    \centering
    \includegraphics[width=0.495\linewidth]{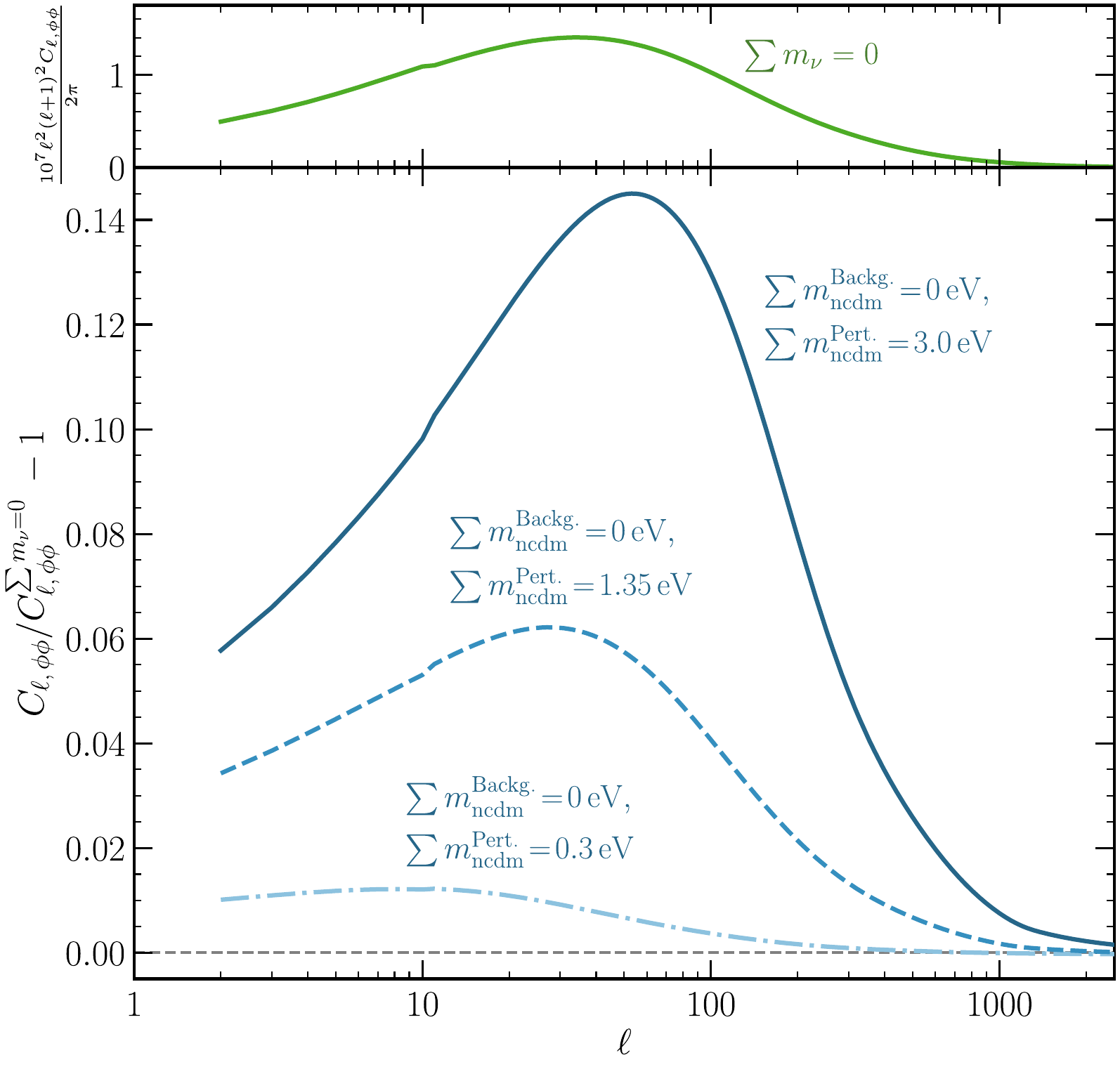}
    \caption{Scale-dependent impact of $\sum m_\nu^{\mathrm{Pert.}}$ on the CMB lensing power spectrum. $\sum m_\nu^{\mathrm{Pert.}}$ increases gravitational potentials through neutrino clustering, increasing CMB lensing. Neutrino clustering is a scale-dependent effect, controlled by the scale $k_{\mathrm{FS}}$ (see~\cref{eq:k_FS}). \textit{A larger $\sum m_\nu^{\mathrm{Pert.}}$ increases $k_{\mathrm{FS}}$, shifting the impact of neutrino clustering to higher multipoles.}}
    \label{fig:cmb-lensing-kFS-scale}
\end{figure}

As mentioned above, the impact of $\sum m_\nu^{\mathrm{Pert.}}$ is scale-dependent, with the characteristic scale $k_{\mathrm{FS}}$ being proportional to $\sum m_\nu^{\mathrm{Pert.}}$ (see \cref{eq:k_FS}). \Cref{fig:cmb-lensing-kFS-scale} shows this effect and how $\sum m_\nu^{\mathrm{Pert.}}$ controls it. We plot the power spectrum of the lensing potential, which is directly related to gravitational potentials~\cite{Dodelson:2020abc,Lewis:2006fu} (the scale dependence in CMB anisotropies is less evident, because lensing is non-linear and the translation between $k$ and $\ell$ is not straightforward). As the figure shows, $\sum m_\nu^{\mathrm{Pert.}}$ enhances lensing, and increasing $\sum m_\nu^{\mathrm{Pert.}}$ shifts its strongest impact to higher $\ell$. 

\begin{figure}[b]
    \centering
    \includegraphics[width=0.495\linewidth]{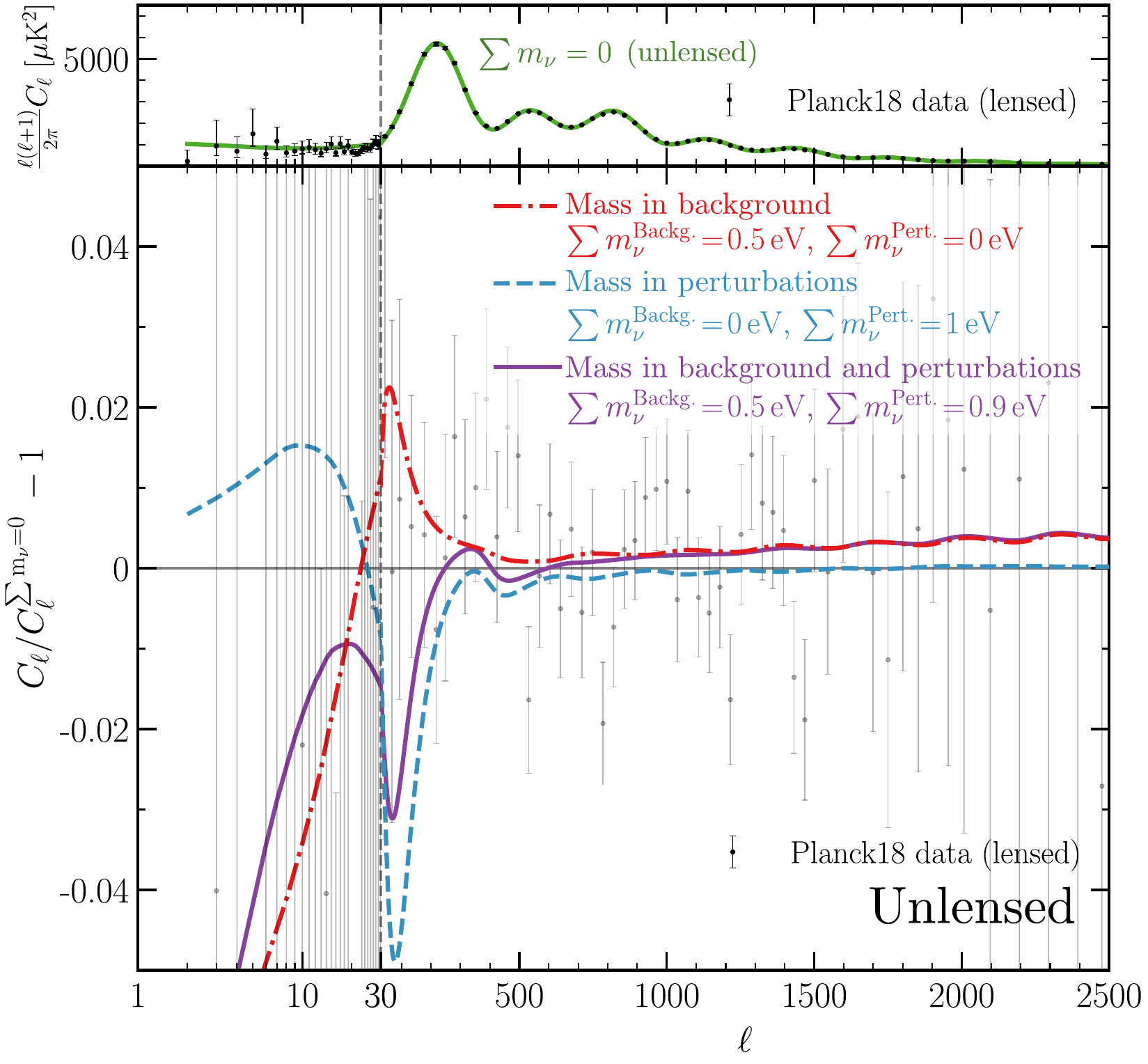}
    \caption{Impact on CMB anisotropies of parameters excluded by our analysis, without gravitational lensing. This Figure is analogous to~\cref{fig:cmb2} in the main text. 
    Since the predictions are unlensed, but data is lensed, here the Planck18 data only shows visually the uncertainty of the measurements. 
    \textit{CMB bounds on $\sum m_\nu^{\mathrm{Backg.}}$ are mainly due to the effect of lensing at high-$\ell$.}}
    \label{fig:cmb-2-unlensed}
\end{figure}

Finally, for completeness and to understand the relevance of lensing in setting CMB limits, \cref{fig:cmb-2-unlensed} shows the predictions for~\cref{fig:cmb2} without lensing. A value of $\sum m_\nu^{\mathrm{Backg.}}$ that is excluded by our analysis predicts an unlensed damping tail well within error bars. Therefore, the constraining power of Planck 2018 data on $\sum m_\nu^{\mathrm{Backg.}}$ at high $\ell$ comes from its effect on CMB lensing. 

\clearpage
\newpage
\section{Full results of the statistical analysis}
\label{sec:app_full_analysis}

In this Appendix, we provide the full results of our statistical analysis. We carry out a Bayesian analysis with the Markov Chain Monte Carlo code \texttt{COBAYA}~\cite{Torrado:2020dgo, 2019ascl.soft10019T}, as mentioned in the main text.

\Cref{tab:priors} contains the priors on the cosmological parameters over which we scan.

\begin{table}[h!t]
\centering
\setlength{\tabcolsep}{10pt}
\renewcommand{\arraystretch}{1.5}
\resizebox{\textwidth}{!}{
\begin{tabular} {c|c c c c c c c c}
Parameter & {$\log(10^{10} A_\mathrm{s})$} & {$n_\mathrm{s}   $} & {$100\theta_\mathrm{s}$} & {$\omega_\mathrm{b}$} & {$\omega_\mathrm{cdm}$} & {$\tau_\mathrm{reio}$} & {$\sum m_\nu^\mathrm{Backg.}~\mathrm{[eV]}$} & {$\sum m_\nu^\mathrm{Pert.}~\mathrm{[eV]}$} \\
\midrule 
Prior & $\mathcal{U}[1.61, 3.91]$ & $\mathcal{U}[0.8, 1.2]$ & $\mathcal{U}[0.5, 10]$ & $\mathcal{U}[0.005, 0.1]$ & $\mathcal{U}[0.001, 0.99]$ & $\mathcal{U}[0.01, 0.8]$ & $\mathcal{U}[0.0, 3.0]$  & $\mathcal{U}[0.0, 3.0]$ \\ 
\end{tabular}}
\caption{Cosmological parameters that we scan over, and their corresponding priors. $\mathcal{U}[a,b]$ denotes a uniform distribution with lower limit $a$ and upper limit $b$.} 
\label{tab:priors}
\end{table}

\begin{figure}[b]
    \centering

    \includegraphics[width=\columnwidth]{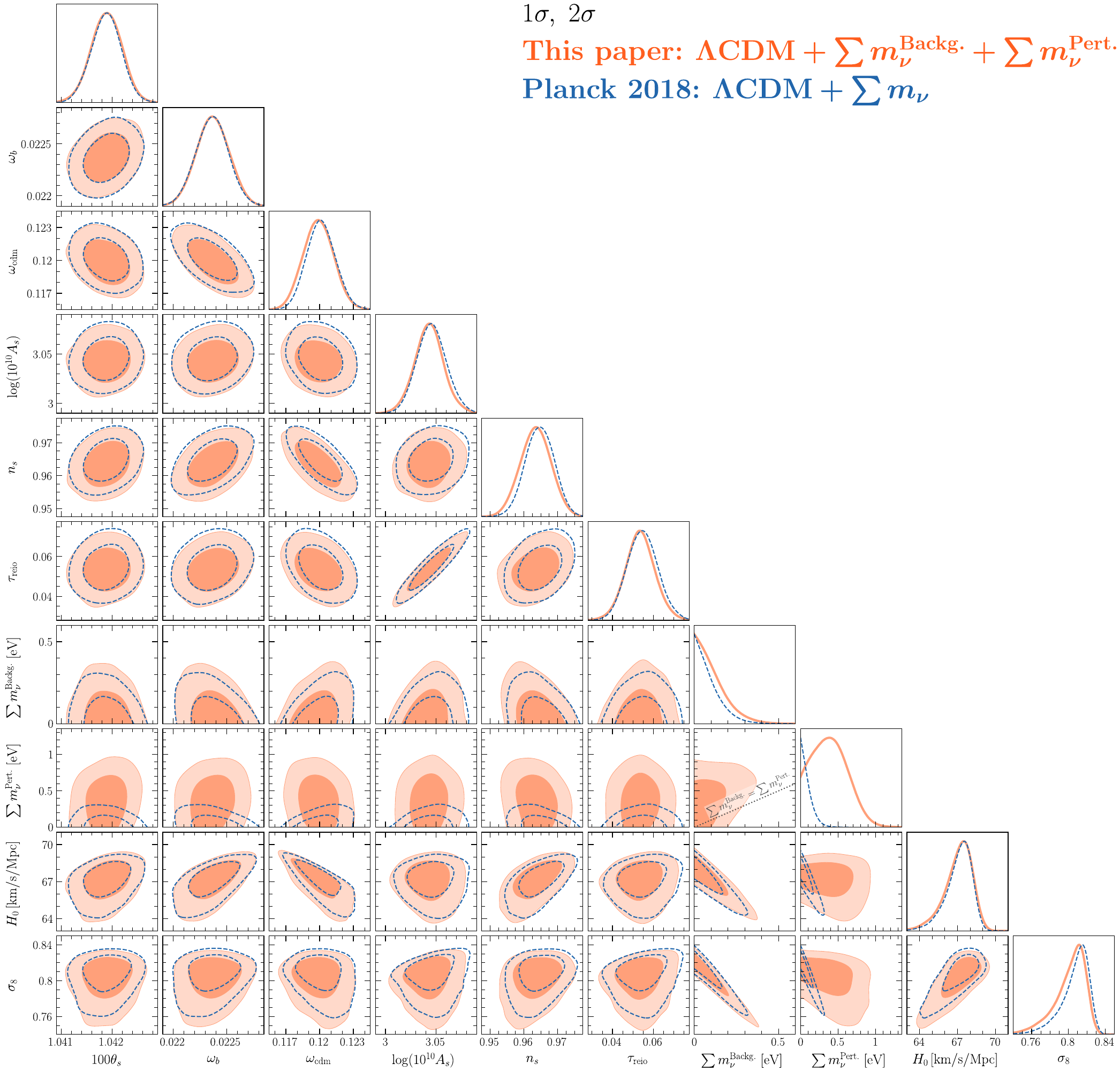}
    
\caption{CMB $1\sigma$ and $2\sigma$ credible regions for all parameters in our analysis, together with individual posterior probabilities. Dashed lines correspond to the standard results, where neutrino-mass effects are not split between background and perturbations. In each subfigure, unshown parameters are marginalized over.}

    \label{fig:triangle_total}
\end{figure}

\Cref{fig:triangle_total} shows the 1D posterior probabilities and 2D credible regions for all parameters in our analysis. We also include the posteriors on the derived parameters $H_0$ and $\sigma_8$, as they show the largest degeneracies with neutrino masses. All posteriors are well-contained within their priors, and the convergence of the MCMC run is determined by an $R-1<0.02$ Gelman-Rubin test~\cite{Lewis:2013hha,10.1214/ss/1177011136}.

\Cref{tab:resultCOB} summarizes the $1$--$3\sigma$ credible intervals on the total neutrino mass, both for the standard scenario ($\sum m_\nu^\mathrm{Backg.} = \sum m_\nu^\mathrm{Pert.}$) and for our analysis that splits among background and perturbations neutrino-mass effects.

\begingroup
\setlength{\tabcolsep}{10pt}
\renewcommand{\arraystretch}{1.8}
\begin{table}[hb]
    \centering
    \textbf{Planck 2018; TT, TE, EE+lowE+lensing}
    
    \begin{tabular}{cccc}
         \toprule
        Mass type& $68\%$ {CL} & {$95\%$} {CL} & {$99\%$} {CL} \\
        \midrule
        \color{Planck18} $\boldsymbol{\sum m_\nu}$ & $<0.11\,\mathrm{eV}$ & $<0.24\,\mathrm{eV}$ & $<0.35\,\mathrm{eV}$ \\
        \color{Thispaper}$\boldsymbol{\sum m_\nu^\mathrm{Backg.}}$ & $<0.13\,\mathrm{eV}$ & $<0.29\,\mathrm{eV}$ & $<0.40\,\mathrm{eV}$ \\
        \color{Thispaper}$\boldsymbol{\sum m_\nu^\mathrm{Pert.}}$ & $0.40^{+0.19}_{-0.29}\,\mathrm{eV}$ & $<0.79\,\mathrm{eV}$ & $<0.97\,\mathrm{eV}$ \\
        \bottomrule
    \end{tabular}
    \caption{Neutrino-mass limits at different confidence levels. These results correspond to the posteriors shown in \cref{fig:triangle_total}. $\sum m_\nu$ refers to the standard scenario in which $\sum m_\nu^\mathrm{Backg.} = \sum m_\nu^\mathrm{Pert.}$.}
    \label{tab:resultCOB}
\end{table}
\endgroup

Both in \cref{fig:triangle_total,tab:resultCOB}, there is a high degree of consistency when comparing $\sum m_\nu^\mathrm{Backg.}$ with Planck 2018 results on $\sum m_\nu$. This is one of our main results, i.e., that the CMB is more sensitive to the temporal dilution of the neutrino energy density than to the ``kinematic'' effects of neutrino masses. The limits on $\sum m_\nu^\mathrm{Backg.}$ are slightly weaker than those on $\sum m_\nu$ due to the degeneracy with $\sum m_\nu^\mathrm{Pert.}$ discussed in the main text. 

\begin{figure}[b]
    \centering

    \includegraphics[width=\columnwidth]{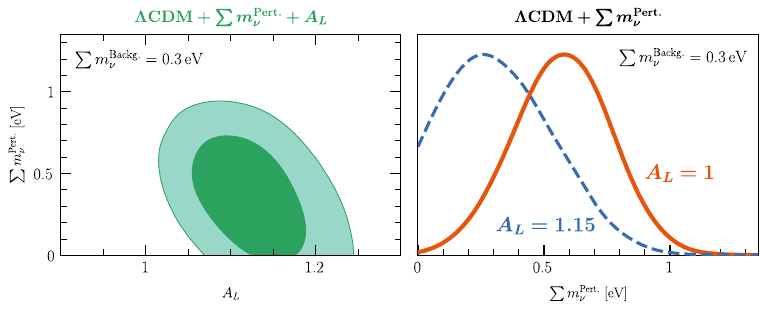}
    
\caption{\textit{Left:} CMB $1\sigma$ and $2\sigma$ credible regions for $\sum m_\nu^\mathrm{Pert.}$ and $A_L$, fixing $\sum m_\nu^\mathrm{Backg.}=0.3\,\mathrm{eV}$. There is a clear correlation between both parameters: for higher $A_L$, the preferred value of $\sum m_\nu^\mathrm{Pert.}$ decreases. \textit{Right:} 1D posterior probabilities for $\sum m_\nu^\mathrm{Pert.}$, fixing $\sum m_\nu^\mathrm{Backg.}=0.3\,\mathrm{eV}$ and $A_L$. For larger $A_L$, the distribution shifts towards $\sum m_\nu^\mathrm{Pert.} = 0$.}
    \label{fig:ALcheck}
\end{figure}

\begin{figure}[b]
    \centering

    \includegraphics[width=0.57\columnwidth]{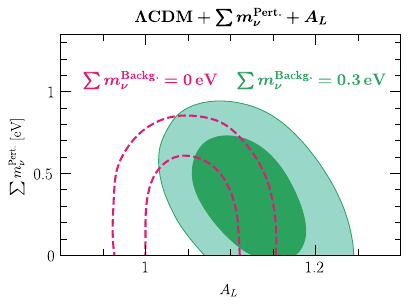}
    
\caption{CMB $1\sigma$ and $2\sigma$ credible regions for $\sum m_\nu^\mathrm{Pert.}$ and $A_L$, fixing $\sum m_\nu^\mathrm{Backg.}$. When $\sum m_\nu^\mathrm{Backg.} = 0$, the correlation among $\sum m_\nu^\mathrm{Pert.}$ and $A_L$ disappears. Hence, when marginalizing over $\sum m_\nu^\mathrm{Backg.}$, the preference for $\sum m_\nu^\mathrm{Pert.} \neq 0$ when $A_L=1$ gets diluted.}

    \label{fig:AL_BCK_check}
\end{figure}

We also note from \cref{fig:triangle,fig:triangle_total,tab:resultCOB} that the best fit for $\sum m_\nu^\mathrm{Pert.}$ is non-zero, with $\sum m_\nu^\mathrm{Pert.} = 0$ being excluded at $\sim 1\sigma$. This is driven by the lensing anomaly ($\sum m_\nu^\mathrm{Pert.}$ increases CMB lensing, see \cref{sec:phenomenology} in the main text and \cref{sec:app_lensing} above), but the result is not statistically significant. The effect is further illustrated in \cref{fig:ALcheck,fig:AL_BCK_check}. There, we repeat our analysis with an additional phenomenological parameter, $A_L$, that rescales the overall amplitude of CMB lensing.

The left panel of \cref{fig:ALcheck} shows the 2D credible regions for $\sum m_\nu^\mathrm{Pert.}$ and $A_L$, with fixed $\sum m_\nu^\mathrm{Backg.}=0.3\,\mathrm{eV}$. The degeneracy between non-zero $\sum m_\nu^\mathrm{Pert.}$ and enhanced CMB lensing is clear. The same effect is visible in the right panel, that shows the 1D posterior on $\sum m_\nu^\mathrm{Pert.}$ for $A_L=1$ (i.e., standard CMB lensing) and $A_L=1.15$ (i.e., enhanced CMB lensing). If $A_L=1$ there is a strong preference for $\sum m_\nu^\mathrm{Pert.} \neq 0$. This preference is driven by the lensing anomaly, and it gets consequently diluted when $A_L = 1.15$.

In our full analysis, the preference for $\sum m_\nu^\mathrm{Pert.} \neq 0$ is not as strong as the $A_L = 1$ case in the right panel of \cref{fig:ALcheck}. This is because our full analysis marginalizes over $\sum m_\nu^\mathrm{Backg.}$, and $\sum m_\nu^\mathrm{Pert.}$ is only degenerate with enhanced lensing for large enough $\sum m_\nu^\mathrm{Backg.}$ (otherwise, the neutrino energy density gets strongly diluted and the impact of perturbations effects is minor). This is quantified in \cref{fig:AL_BCK_check}, that shows the 2D credible regions for $\sum m_\nu^\mathrm{Pert.}$ and $A_L$ for fixed $\sum m_\nu^\mathrm{Backg.}=0$. As the figure shows, when $\sum m_\nu^\mathrm{Backg.}=0$ there is no degeneracy between $\sum m_\nu^\mathrm{Pert.}$ and enhanced CMB lensing.

\end{document}